\documentstyle[emulateapj,epsfig]{article}
\slugcomment{03 Nov 1999 version, submitted to ApJ}

\def\Vo{\ifmmode V_0\else $V_0$\fi}
\def\VCO{\ifmmode {\rm V_{CO}}\else V$_{\rm CO}$\fi}
\def\Ho{\ifmmode H_0\else $H_0$\fi}
\def\HI{{\ion{H}{1}}}
\def\HII{{\ion{H}{2}}}
\def\H2{\ifmmode {\rm H_2}\else H$_2$\fi}
\def\Ha{\ifmmode H\alpha\else H$\alpha$\fi}
\def\Hb{\ifmmode H\beta\else H$\beta$\fi}
\def\LB{\ifmmode L_B\else $L_B$\fi}
\def\LIR{\ifmmode L_{\rm IR}\else $L_{\rm IR}$\fi}
\def\LHa{\ifmmode L_{\rm H\alpha}\else $L_{\rm H\alpha}$\fi}
\def\LCO{\ifmmode L_{\rm CO}\else $L_{\rm CO}$\fi}
\def\MHI{\ifmmode M_{\rm HI}\else $M_{\rm HI}$\fi}
\def\MH2{\ifmmode M_{\rm H_2}\else $M_{\rm H_2}$\fi}
\def\ICO{\ifmmode I_{\rm CO}\else $I_{\rm CO}$\fi}
\def\SCO{\ifmmode S_{\rm CO}\else $S_{\rm CO}$\fi}
\def\SHI{\ifmmode S_{\rm HI}\else $S_{\rm HI}$\fi}
\def\FHa{\ifmmode F_{\rm H\alpha}\else $F_{\rm H\alpha}$\fi}
\def\FIR{\ifmmode F_{\rm IR}\else $F_{\rm IR}$\fi}
\def\ftwelve{\ifmmode S_{12}\else $S_{12}$\fi}
\def\ftwenty{\ifmmode S_{25}\else $S_{25}$\fi}
\def\fsixty{\ifmmode S_{60}\else $S_{60}$\fi}
\def\fhundred{\ifmmode S_{100}\else $S_{100}$\fi}
\def\Tdust{\ifmmode T_{\rm dust}\else $T_{\rm dust}$\fi}
\def\WHa{\ifmmode W_{\lambda}(\Ha)\else $W_{\lambda}$(\Ha)\fi}
\def\Lsun{\ifmmode L_{\sun}\else $L_{\sun}$\fi}
\def\Msun{\ifmmode M_{\sun}\else $M_{\sun}$\fi}
\def\IRAS{{\it IRAS\/}}
\def\Bt{\ifmmode B_{\rm T}\else $B_{\rm T}$\fi}
\def\Ab{\ifmmode A_B\else $A_B$\fi}
\def\Av{\ifmmode A_V\else $A_V$\fi}
\def\kms{\ifmmode km~s^{-1}\else km~s$^{-1}$\fi}
\def\um{\ifmmode \mu m\else $\mu$m\fi}
\def\lt{$<$}
\def\gt{$>$}
\def\ap{$\sim$}
\def\mp{$\pm$}
\def\nodd{\phd\nodata}
\lefthead{Bushouse, Lord, Lamb, Werner, \& Lo}
\righthead{Molecular Gas in Interacting Galaxies}

\begin{document}

\title{The Molecular Gas---Star Formation Connection in an Optically-Selected
 Sample of Interacting Galaxies}

\author{Howard~A.~Bushouse\altaffilmark{1},
 Steven~D.~Lord\altaffilmark{2},
 Susan~A.~Lamb\altaffilmark{3},
 Michael~W.~Werner\altaffilmark{4}, and
 K.~Y.~Lo\altaffilmark{5,6}
}

\altaffiltext{1}{Space Telescope Science Institute, 3700 San Martin Dr.,
 Baltimore, MD 21218; bushouse@stsci.edu}
\altaffiltext{2}{Infrared Processing and Analysis Center, California Institute
 of Technology, M.S. 100-22, Pasadena, CA 91125; lord@ipac.caltech.edu}
\altaffiltext{3}{Department of Physics, University of Illinois, 1110 W.~Green
 St., Urbana, IL 61801; slamb@astro.uiuc.edu}
\altaffiltext{4}{Jet Propulsion Laboratory, California Institute of Technology,
 M.S. 264-767, 4800 Oak Grove Dr., Pasadena, CA 91109; mww@ipac.caltech.edu}
\altaffiltext{5}{Department of Astronomy, University of Illinois, 1002 W.~Green
 St., Urbana, IL 61801; kyl@astro.uiuc.edu}
\altaffiltext{6}{Current address: Academia Sinica Institute of Astronomy and
 Astrophysics, P.O. Box 1-87, Nankang, Taipei, Taiwan 115, R.O.C.;
 kyl@asiaa.sinica.edu.tw}

\begin{abstract}
We have obtained $^{12}$CO (1--0) emission-line observations for
a sample of 37 interacting galaxy systems, chosen from a parent sample of
optically-selected interacting galaxies.
The sample observed here spans a large range of interaction strengths and
current star formation rates.
Using the standard Galactic CO-to-\H2\ conversion factor we find that the
interacting galaxies are, on average, marginally more rich in molecular gas
than a comparison sample of isolated spiral galaxies, having mean
\MH2/\LB\ and \MH2/\MHI\ ratios that are 20--40\%\ higher than that of the
isolated spirals.
The interacting galaxies have a mean \LIR/\MH2\ ratio that is a factor
of $\sim$1.3 higher than the isolated spirals, indicating that the rate of
high-mass star formation per unit molecular gas---the star formation
efficiency---is also enhanced within the interacting galaxy sample.
There is a strong correlation between relative \H2\ content and star formation
rates derived from both far-infrared and \Ha\ luminosities, indicating that the
level of interaction-induced star formation activity is highly dependent upon
the available gas supply.
There are some individual galaxies with moderate amounts of molecular gas but
little or no current star formation activity.
Therefore molecular gas is a necessary, but not sufficient, condition for 
interaction-induced star formation activity.
There is a strong correlation between interaction strength and both star
formation rate and fractional \H2\ content, with merging systems
showing the highest star formation rates and fractional \H2\ contents.
There is also an increase in the molecular to atomic gas ratio with
increasing interaction strength.
If the increase in derived \H2\ content is real, there must be a substantial
conversion of \HI\ to \H2\ gas taking place in the strongest interactions.
At least some portion of these trends, however, may arise due to an increase
in cloud heating, brought about by the increased population of young, massive
stars in galaxies with high current star formation rates.
A reduction in the standard CO-to-\H2\ conversion ratio by a factor of two
could account for the systematic increase in derived \H2\ content as a
function of interaction strength.
Such a change could not, however, account for the total range of 
\H2\ contents within the entire samples of galaxies.
Therefore the correlation between molecular gas content and star formation
rate must be real.

\end{abstract}


\section{Introduction}
Studies of the global molecular gas content in interacting galaxy systems have
shown that, on average, these galaxies appear to be rich in molecular gas and
have star formation efficiences---as measured by the formation rate of
high-mass stars per unit molecular gas mass---that are much higher than those
in isolated spiral galaxies (\cite{san86}; \cite{sol88}; \cite{tin90};
\cite{san91}; \cite{sof93}; \cite{you96}).
Furthermore, both molecular gas content and star formation efficiency (SFE)
appear to be correlated with interaction strength, with strongly interacting
and merging systems having the highest molecular gas contents and SFE's.

Most of these studies have focused on samples of galaxies that were chosen
for their high far-infrared fluxes, since the observed correlation
between infrared and CO flux (e.g. Young et~al.~1984, 1986; \cite{san85};
\cite{sol88}) ensures detectability in CO surveys.
The presence of interacting and merging galaxies in these surveys is a
bi-product of the fact that the relative fraction of interacting galaxies
increases with far-IR luminosity (e.g. \cite{soi84}; \cite{kle87};
Sanders et~al.~1987, 1988a, 1988b).

In this study we investigate the molecular gas content of a sample of
interacting systems that were selected according to their optical morphology.
Therefore this study differs from previous investigations of CO emission
in interacting galaxies, which have for the most part concentrated on
infrared-luminous and merger type objects.
The systems that have been studied are part of a larger sample that was
selected on the basis of optical morphological evidence for strong tidal
interactions and has been well studied at optical, infrared, and radio
wavelengths.
The goal of this study is not only to measure the molecular gas
content of interacting galaxies, but more importantly to examine the
relationship between gas content and star formation characteristics, as well
as interaction strength.
Questions that this study will address include whether or not a rich
molecular gas disk is a prerequisite for a starburst in interacting systems
or if enhanced molecular gas production is occuring;
whether optically ``dormant'' pair members have molecular material present;
how the molecular gas content, far-IR luminosity and colors, and optical
indicators of star formation are related in strongly interacting galaxies;
and what the relationship is between molecular and atomic gas content.
By addressing these questions, we hope to achieve a better idea of the
circumstances under which starbursts occur in interactions and thus of the
mechanisms which trigger the burst.

\section {The Samples}

The galaxies for this study have been selected from the large sample
of interacting systems compiled by \cite{bus86}.
Members of this sample were originally chosen on the basis of optical
morphology only, preferentially selecting systems with features
characteristic of strong interactions, such as tidal tails, loops, and bridges.
Hence it is not intentionally biased towards infrared- or radio-bright
systems, nor galaxies containing active galactic nuclei (AGN's), although
many systems in the parent sample do have these characteristics.
Another advantage in choosing systems from this sample is that an
extensive and well-studied database of optical spectra, \Ha\ images, near- and
far-IR photometry and imaging, and \HI\ 21 cm data exists
(Bushouse 1986, 1987; \cite{bus88}; \cite{bus90}; \cite{bus98}).

We have obtained $^{12}$CO (1--0) observations for 24 systems from the sample
of \cite{bus86}.
Some of the observed systems were selected because of their
relatively high far-IR fluxes, which, because of the well-known
correlation between far-IR and CO fluxes for galaxies, suggested that
they would be easily detectable in the CO emission line.
Other systems observed in this study were chosen in an attempt to fully sample
the parameter space of known optical, infrared, and radio characteristics of
the class of interacting galaxies.
For example, while infrared-bright systems often have optical indicators
of high current star formation rates (SFR's), some systems were chosen because
they have optical indications of low SFR's, regardless of their infrared flux
levels.
Others were chosen because they have optical indicators of high current SFR's,
yet have low far-IR emission levels.
Many systems were also chosen because of the relatively large angular
separation of the two galaxies in the pair, allowing separate CO measurements
for each of the galaxies and subsequent analysis of the properties of the
individual galaxies.

We also searched the literature for CO observations of systems contained in
the \cite{bus86} parent sample.
This search yielded a total of 19 observations of galaxies in 13 interacting
systems.
Most of these data are for infrared-bright systems, as previous studies
mainly used IR-selected samples.
The combined sample of interacting systems analyzed in this study is not
complete in either a flux-limited or volume-limited sense; rather, the
galaxies were selected to span a wide range of parameter space, including
far-IR flux level and optical indications of star formation properties.

Table~\ref{tbl_codata} contains the list of interacting systems from the sample
of \cite{bus86} for which CO observations have been obtained.
It also contains a comment on the nature of each system and the
CO observations that have been obtained.
There are a total of 19 paired systems, refered to as ``complete pairs'', for
which individual CO observations of the two galaxies in each pair are available.
There are another 5 systems for which data for only one of the two
galaxies (``1 of 2'') in the pair has been obtained.
Finally, there are 13 systems that have a single CO observation that
encompasses the entire pair.
Nine of these are compact systems in which the pair of galaxies fits within a
single CO aperture.
The remaining four systems---NGC~1614, UGC~966, UGC~4509, and UGC~8387---are
generally accepted to be systems that are in an advanced stage of merging,
where the two original galaxies are in the process of coalescing into a single
object.
CO emission has been detected in 41 of the 56 observations.

We have also used a sample of isolated spiral galaxies from the Five College
Radio Astronomy Observatory (FCRAO) Extragalactic CO Survey (Young et~al.~1995)
as a comparison sample in our study of the interacting systems.
We selected 80 galaxies from this survey, ranging in morphological type from
S0 to Sd, intentionally excluding irregulars and known interacting systems.
The galaxies were chosen simply on the basis of available \IRAS\ and CO
observations.

\section{Observations}

The $^{12}$CO (1--0) observations were obtained in December 1988 using the
12~m NRAO telescope\footnotemark\ (HPBW = 55\arcsec\ at 115 GHz) on Kitt Peak.
\footnotetext{The National Radio Astronomy Observatory is operated by Associated
Universities, Inc., under contract with the National Science Foundation.}
The telescope was equipped with dual polarization SIS receivers
($T_{ssb} \sim 110$ K).
Two 256~x~2 MHz channel filter banks, one for each polarization, provided
a total velocity coverage of approximately 1360 \kms\ with a resolution of
$\sim$5.6 \kms\ per channel.
Telescope pointing was monitored by observations of planets and was
estimated to be accurate to \mp4\arcsec (rms).
All of the observations were taken using the nutating subreflector, which gave
very flat baselines.
The spectra were calibrated using an ambient temperature chopper wheel
and scaled by the telescope efficiency on a spatially extended source,
$\eta_{fss}$ = $\eta_{Moon}$ to give line temperatures $T_{R}^{*}$.
After co-adding both polarizations, the CO spectra were Gaussian smoothed
to a resolution of 15--20 \kms, giving an average rms noise of 3--4 mK per
channel.

The CO spectra were obtained with the spectrometer bandpass centered on the
known \HI\ or optical redshift.
In cases of weak detections, the center of the spectrometer bandpass was
shifted by a few hundred \kms\ while observing a given target, in order to
confirm the reality of the detection.
Observations were obtained for a single pointing at the optical coordinates
of the galaxies.
For systems in which the two galaxies of the pair easily fit within the
telescope half-power beam width (55\arcsec), the telescope was pointed midway
between the two galaxies.
Most of the systems that were observed for this study are sufficiently distant
that the total CO flux of the galaxies has been measured within a single
telescope beam.
We have not attempted to map any of the systems, nor apply any corrections to
the observed fluxes for emission that might fall outside the beam area.
Some of the data for interacting systems taken from the literature, however,
are for more nearby systems with correspondingly larger angular sizes.
The total line fluxes for many of these are the result of mapping observations.

The final co-added, baseline subtracted, smoothed spectrum for each detected
source is shown in Fig.~\ref{fig_cospectra}.
The total CO intensity for each source was obtained by integrating the
spectrum between velocity limits encompassing the observed line.
Multiple independent flux measurements from the spectra show a 
1$\sigma$\ scatter of 3\%.
Independent observations of six galaxies within the interacting sample, also
made with the NRAO 12m telescope (\cite{zhu99}), yield a mean ratio in measured
fluxes of 0.99, with a 1$\sigma$\ scatter from object to object of 20\%.
We therefore estimate an uncertainty of 20\% in the integrated CO flux for
individual objects.

The integrated CO line fluxes for the interacting galaxies, converted from
\ICO\ (K \kms) to \SCO\ (Jy \kms) using a gain factor of 35 Jy~K$^{-1}
(T_{R}^{*})$ for the NRAO 12 m telescope, are given in Table~\ref{tbl_codata}.
Fluxes taken from the literature have been converted to the \SCO\ scale
using the gain factor appropriate for the telescope that was used to obtain
the data and are also shown in Table~\ref{tbl_codata}.
The measured velocity centers and widths of the CO lines are also given.

A compilation of observed optical, \IRAS, and \HI\ data for the interacting
systems included in this study is given in Table~\ref{tbl_obsdata}.
\IRAS\ fluxes have been measured using the Infrared Processing and Analysis
Center's one-dimensional coadding routine ``SCANPI.''
\HI\ data are from a variety of sources, as noted in the table.
Given the generally large beam sizes of the \HI\ observations, most of these
measurements are totals for each interacting system.
\Ha\ fluxes are from our own spectral or narrow-band imaging observations
(\cite{bus87}) or from the narrow-band imaging observations of
Young et~al.~(1996).
The \Ha\ imaging observations, which generally have a field of view
of a few arc minutes, usually provide a global measurement for the galaxies
in this sample, while the spectral observations used a 22\arcsec\ diameter
aperture and do not always sample an entire galaxy.
Therefore the spectral data provide a lower limit on the total \Ha\ emission.
All of the \IRAS\ SCANPI measurements are global values for each system,
therefore there is only one entry in Table~\ref{tbl_obsdata} for each system.
The \HI\ and \Ha\ measurements that are totals for a pair are also listed with
only one entry per system.

Table~\ref{tbl_derdata} presents derived parameters for the interacting galaxy
sample.
As with the observed data in Table~\ref{tbl_obsdata}, systems that have a global
measurement for a given parameter are listed with a single entry.
The far-IR luminosities, \LIR, were computed from the \IRAS\ 60 and
100 \um\ flux densities, using the procedure outlined in Appendix B of
{\it Catalogued Galaxies and Quarsars Observed in the IRAS Survey, Version 2}
(\cite{ful89}).
The far-IR luminosity, in solar units, is given by
$\LIR=3.94\times 10^{5} D^{2} [2.58\fsixty\ + \fhundred]$,
where $D$ is the distance in Mpc, and \fsixty\ and \fhundred\ are the
\IRAS\ flux densities listed in Table~\ref{tbl_obsdata}.

\H2\ masses were derived from the CO (1--0) fluxes using a constant conversion
factor of 
$N(\H2)/\ICO=2.8\times 10^{20}$ \H2\ cm$^{-2}$ [K(T$_R$) km s$^{-1}$]$^{-1}$
(\cite{blo86}).
Kenney \& Young (1989) show that this value of the conversion factor leads to
\H2\ masses in solar units given by $M(\H2)=1.1\times 10^4D^2\SCO$,
where D is the distance in Mpc and \SCO\ is the CO flux in Jy km s$^{-1}$.
We note that this conversion factor is derived from observations of Galactic
molecular clouds and may not be universally applicable, especially in the
case of strongly interacting galaxies where the state of the ISM could be
substantially different than in the disk of a quiescent galaxy.
For the sake of simplicity and consistency, the observational results and
conclusions presented in \S4 are based on the use of this constant conversion
factor to derive \H2\ gas masses from observed CO emission.
At the very least, the derived values for \H2\ gas mass can be thought of as
a measure of the amount of warm (and therefore visible) CO gas in the
galaxies.
In \S5 the possibility and ramifications of a variable conversion factor are
discussed.

For the comparison sample of isolated galaxies, observed \IRAS\ and \HI\ data
have been taken from the compilation of Young et~al.~(1989), CO fluxes from
Young et~al.~(1995), and \Ha\ fluxes from Young et~al.~(1996).
All derived properties for the isolated galaxies have been computed in the
same fashion as for the interacting galaxies.

\section{Results}

In the following sections the molecular gas properties of the interacting
systems are examined and compared to those of the sample of isolated spirals.
First, the overall properties of the entire samples of interacting and
isolated galaxies are compared, including an examination of the relationships
between various global properties, such as luminosity, gas content, and star
formation rates.
The analysis is then continued by subdividing the interacting systems into
smaller groups, based on star formation characteristics, pair separation, and
interaction strength.

When analyzing properties of the interacting galaxies care is taken to utilize
data for the individual galaxies within each system, when such are available.
When comparing characteristics that are only available on a global basis
with those for which individual measurements exist, the measurements for the
individual galaxies within a system are combined to form a global value, so
that the comparisons are consistent.
For example, far-IR and \HI\ measurements of the interacting systems typically
include both galaxies in a pair, while CO and \Ha\ observations are available
for many of the individual galaxies.
Therefore when comparing far-IR and CO quantities, the CO measurements for the
two galaxies in a pair are summed to compare with the global far-IR measurement.

For each relationship that is examined a least-squares linear fit to the data
for only the isolated galaxies is computed, which then allows any
unusual trends in the data for the interacting systems to be seen.
All plots of these relationships show the computed fit to the isolated galaxy
data, as well as the correlation coefficient of the fit.

\subsection{Properties of the Overall Samples}

Mean and median properties of the entire isolated and interacting galaxy
samples are listed in Table~\ref{tbl_mnprop}.
Included in this table are dust temperatures derived by
fitting \IRAS\ \fsixty\ and \fhundred\ fluxes with a function of the form
$f_{\nu}\ \propto\ \nu^nB_{\nu}(T_d)$, with $n = 1.0$, which is a common
approximation for the broadband spectral energy distributions of galaxies.
In order to properly take into account the non-detections in the samples, the
mean and median statistics listed in Table~\ref{tbl_mnprop} have been determined
using the IRAF/STSDAS task ``kmestimate,'' which computes the Kaplan-Meier
estimator of a randomly censored distribution (\cite{kap58}).
Both mean and median values have been computed because they each have certain
advantages.
It is possible to estimate an uncertainty in the mean, but not the median.
Knowing the uncertainties allows us to test the statistical significance of
any differences between the samples.
The median, on the other hand, is not as easily biased by a few extreme values,
which are often present in the distributions analyzed here, and
can therefore give a more appropriate representation of the overall sample.

\subsubsection{\H2\ Content}

\begin{figure}
\begin{center}
\epsfig{file=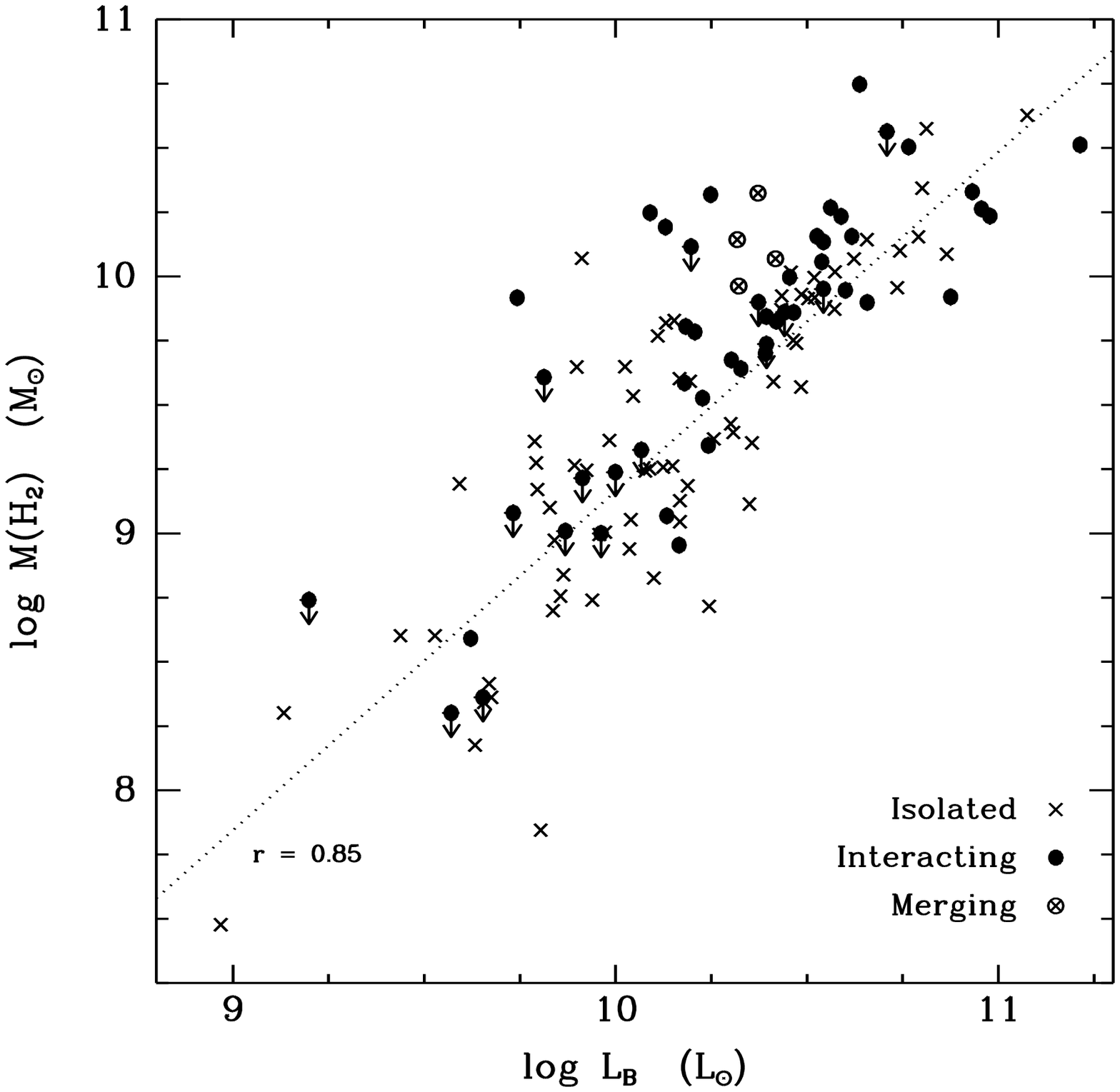,width=6.0in,angle=-90}
\end{center}
\caption{\H2\ gas content as a function of optical blue luminosity for the
isolated and interacting galaxies.
The dotted line is a linear fit to the isolated galaxy data only and $r$ is the
correlation coefficient.
Values for the interacting systems are for individual galaxies, when available,
otherwise they are totals for a system.
\label{fig_lbh2}}
\end{figure}

The \H2\ content of the interacting galaxies, derived from the integrated
CO emission, ranges from $4\times 10^8$ to $5.6\times 10^{10}$ \Msun, with
mean and median values of $9.9\pm1.4 \times 10^9$ \Msun\ and
$6.4 \times 10^9$ \Msun, respectively\footnotemark.
\footnotetext{All quoted errors are the 1$\sigma$ uncertainty in the mean.}
There is complete overlap in the derived \H2\ gas mass values for the
interacting and isolated galaxy samples, but the mean and median values for
the interacting sample are factors of 2--3 times higher than those for the
isolated galaxies (see Table~\ref{tbl_mnprop}).
The larger values for the interacting sample, however, can in large part be
attributed to observational factors.
First, approximately 25\%\ of the interacting galaxy CO observations are of
compact or merging systems that include emission from more than one galaxy.
Second, the interacting galaxy sample as a whole is biased towards more
distant, and hence more luminous, galaxies;
the blue luminosity, \LB, of the interacting sample is 1.5--2 times higher
than that of the isolated galaxy sample.
Fig.~\ref{fig_lbh2} shows that there is a strong correlation between \LB\ and
\MH2, therefore the overall increase in \MH2\ for the interacting sample is
not unexpected.

Furthermore, at the mean distance of the interacting galaxies, our CO
detection limit corresponds to an \H2\ gas mass of $\sim2.5\times 10^9$ \Msun,
which is comparable to the median \MH2\ of the isolated galaxy sample.
Hence the distribution of \MH2\ values for the interacting systems is biased
upwards by our inability to detect low-mass systems.
For example, excluding all Virgo cluster members from the sample of
isolated galaxies---which eliminates most of the nearby systems in that
sample---brings the ratios of mean and median \MH2\ for the interacting and
isolated galaxies down from 2--3 to $\sim$1.5.

\begin{figure}
\begin{center}
\epsfig{file=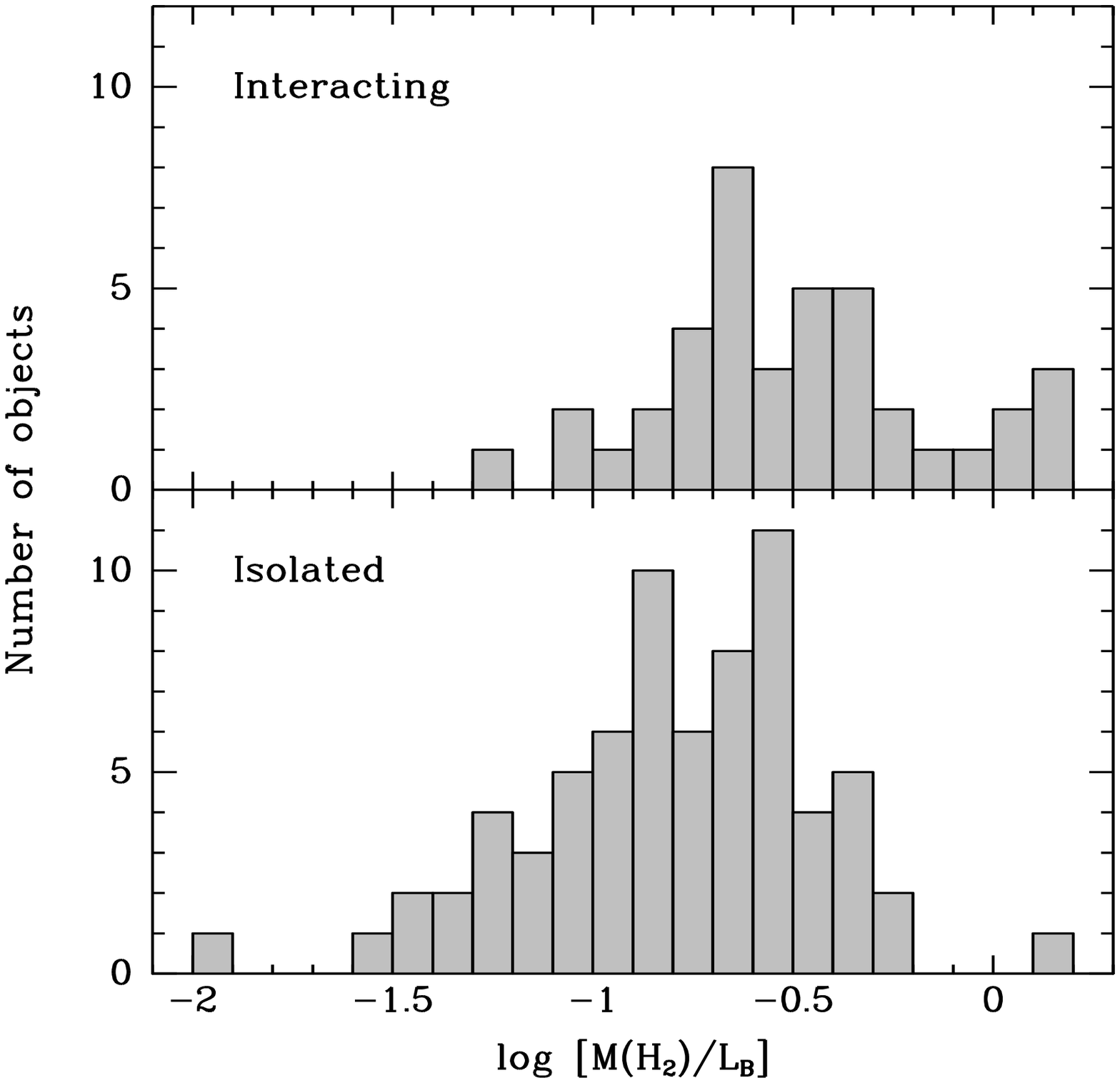,width=6.0in,angle=-90}
\end{center}
\caption{Distributions of \H2\ mass to blue luminosity ratios for the
isolated and interacting galaxies. Values for the interacting systems are for
individual galaxies, when available, otherwise they are averages for a system.
\label{fig_hist_h2lb}}
\end{figure}

Normalizing the \H2\ content by \LB\ helps to remove the biases in the
interacting galaxy sample introduced by sampling more than one galaxy at a
time, as well as the overall increase in galaxy luminosity.
The mean and median \MH2/\LB\ ratios for the interacting systems are factors of
$1.8\pm0.3$ and 1.5 higher, respectively, than those of the isolated galaxies.
Much of this increase, however, can be attributed to our inability to detect
interacting systems with small amounts of \H2\ at the greater distances of
these systems.
This is illustrated graphically in the histograms of \MH2/\LB\ ratios for the
two samples shown in Fig.~\ref{fig_hist_h2lb}.
It can be seen from this figure that much of the increase in both mean and
median \MH2/\LB\ for the interacting systems is due to a lack of systems
with low \MH2/\LB\ values.
Once again, excluding Virgo members from the isolated galaxy sample results in
a drop in the mean and median ratios of \MH2/\LB\ for the two samples to 
values of $1.4\pm0.3$ and 1.0, respectively.
The remaining increase in mean \MH2/\LB\ mainly arises from a small
number of interacting systems that have unusually high \MH2/\LB\ ratios (see
Fig.~\ref{fig_hist_h2lb}), several of which are the advanced merger systems
that were identified in \S2.
Therefore there is only a marginal increase, on average, in fractional
\H2\ content for the interacting systems, with a small number of systems
showing a significant increase.

\begin{figure}
\begin{center}
\epsfig{file=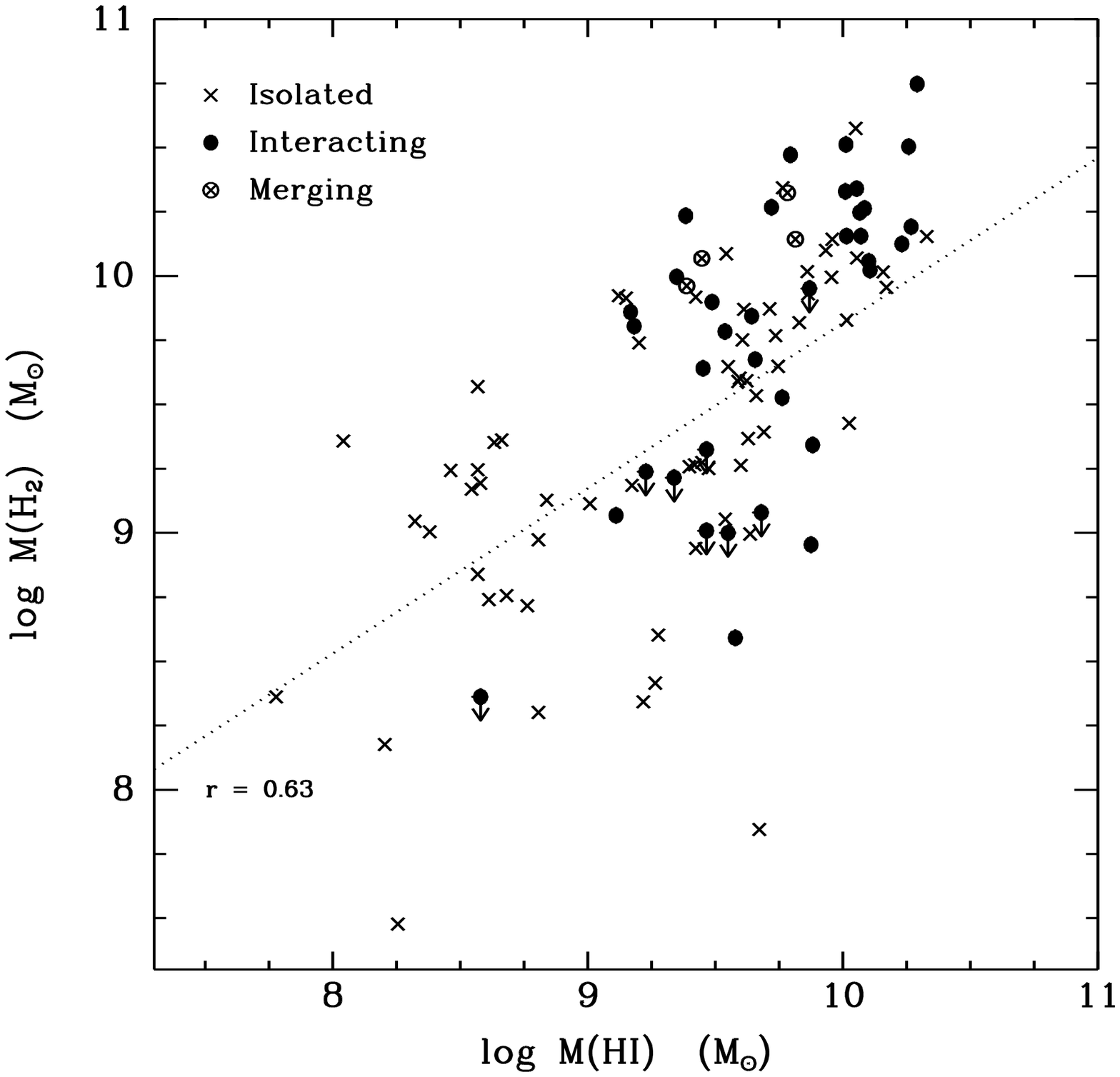,width=6.0in,angle=-90}
\end{center}
\caption{\H2\ gas content as a function of \HI\ gas content for the isolated
and interacting galaxies.
The dotted line is a linear fit to the isolated galaxy data only and $r$ is the
correlation coefficient.
Values for the interacting systems are for individual galaxies, when available,
otherwise they are totals for a system.
\label{fig_hih2}}
\end{figure}

\begin{figure}
\begin{center}
\epsfig{file=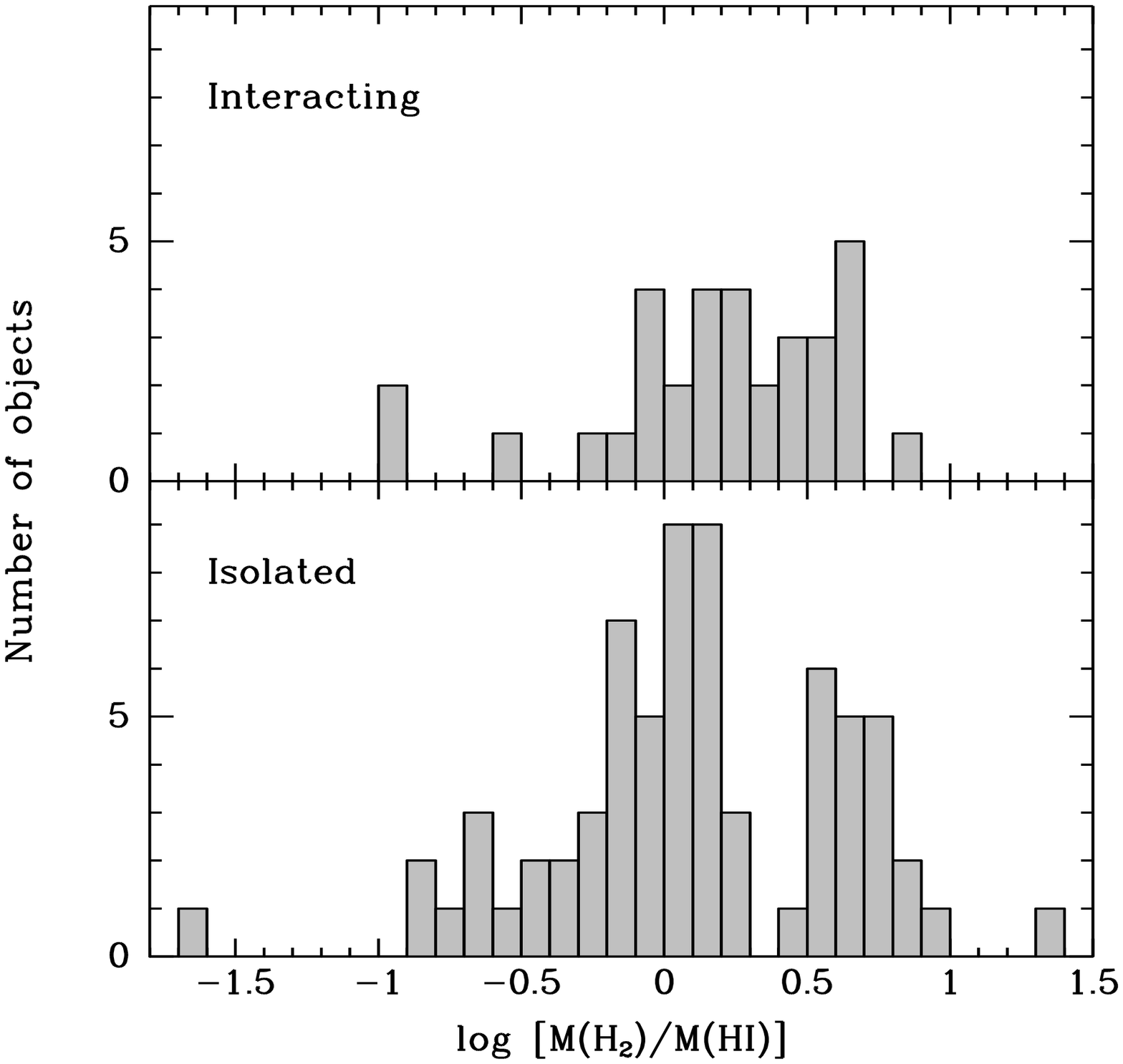,width=6.0in,angle=-90}
\end{center}
\caption{Distributions of \H2\ to \HI\ mass ratios for the isolated and
interacting galaxies. Values for the interacting systems are for individual
galaxies, when available, otherwise they are averages for a system.
\label{fig_hist_h2hi}}
\end{figure}

The correlation shown earlier between \H2\ content and \LB\ (see
Fig.~\ref{fig_lbh2}) indicates that, to first order, the amount of molecular
gas is simply proportional to the overall size of a galaxy.
Figure~\ref{fig_hih2} shows that the total \H2\ and \HI\ contents are
similarly related.
The correlation between \H2\ and \HI, however, shows a larger scatter,
especially for low-mass galaxies, where the molecular gas content can
vary by almost two orders of magnitude for a given amount of atomic gas.
The interacting and isolated galaxy samples have comparable mean
\H2/\HI\ ratios, while the median ratios indicate an increase of $\sim$40\%\ in
fractional \H2\ content for the interacting systems.
Histograms of the \MH2/\MHI\ ratios confirm that there is a similar
distribution for the two samples (Fig.~\ref{fig_hist_h2hi}).
The increase in median \MH2/\MHI\ for the interacting galaxies again appears
to be mainly due to a relative lack of systems with low values, rather than
the presence of systems with higher than normal values.
Note, however, that exclusion of the Virgo cluster members from the sample of
isolated galaxies significantly reduces the mean \MH2/\MHI\ ratio for that
sample, due to the fact that the Virgo galaxies are systematically deficient
in \HI.
Without the Virgo galaxies, the sample of isolated galaxies has a mean
\MH2/\MHI\ ratio that is a factor of $1.2\pm0.3$ lower than that of the
interacting galaxies.

\subsubsection{Star Formation Rate Versus \H2\ Content}

It is well established that the far-IR luminosity of a galaxy can be a
good measure of its current SFR (see \cite{ken98} and references therein).
Young, high-mass stars heat the interstellar dust within and around star
forming regions and the warm dust then reradiates the stellar optical and UV
light into the thermal infrared.
\Ha\ emission provides another measure of the current formation rate
of high-mass stars (\cite{ken98}).
So in order to investigate the relationship between current star formation
rate and molecular gas content in these galaxies, a comparison of their
far-IR and \Ha\ emission levels with derived \H2\ content has been performed.

\begin{figure}
\begin{center}
\epsfig{file=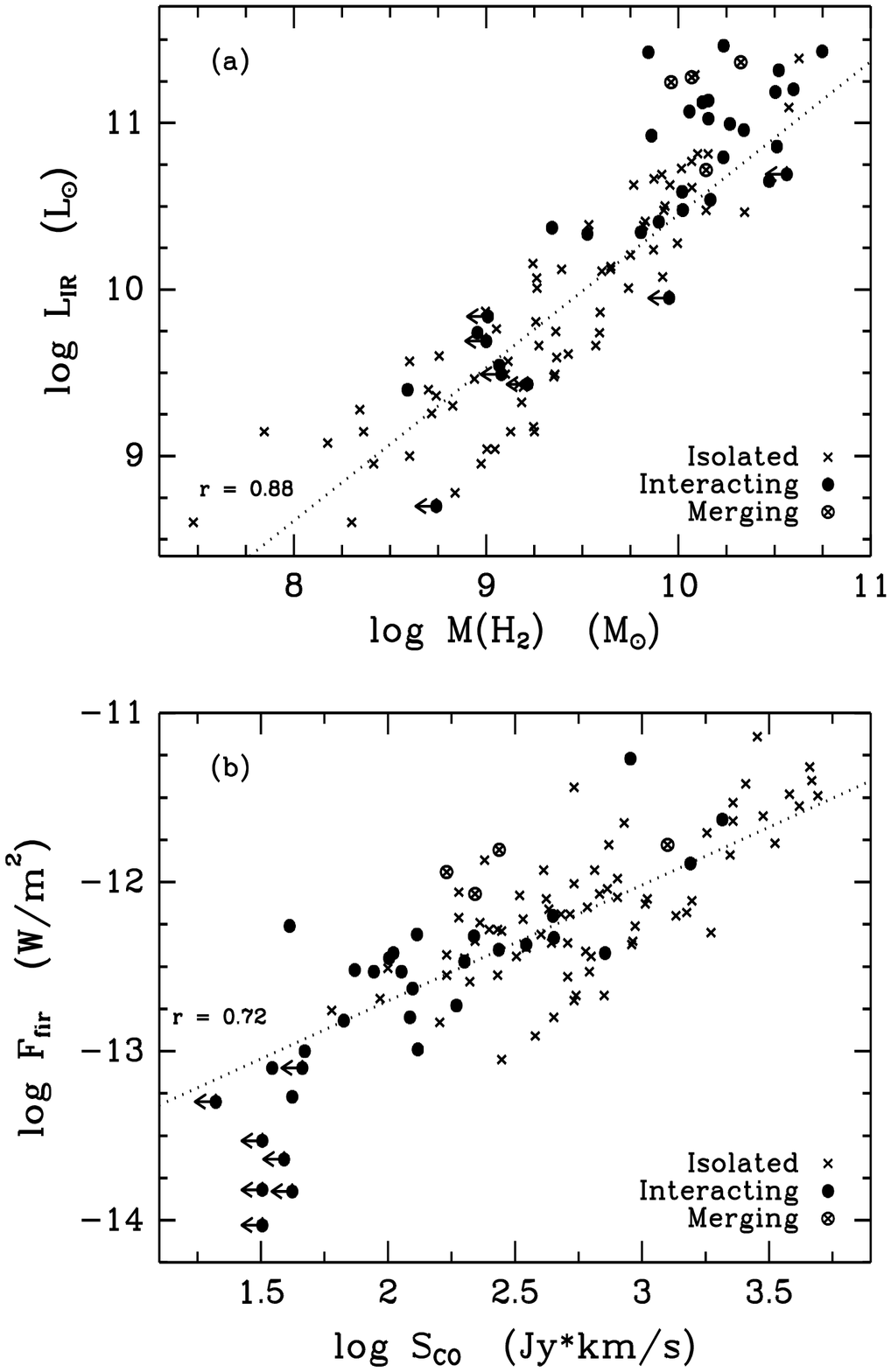,width=5.5in,angle=0}
\end{center}
\caption{(a) Far-infrared luminosity versus \H2\ content and (b)
far-infrared versus CO flux for the isolated and interacting galaxies.
The dotted lines are linear fits to the isolated galaxy data only and $r$ is the
correlation coefficient.
All values for the interacting galaxies are totals for a system.
\label{fig_ir_co}}
\end{figure}

\begin{figure}
\begin{center}
\epsfig{file=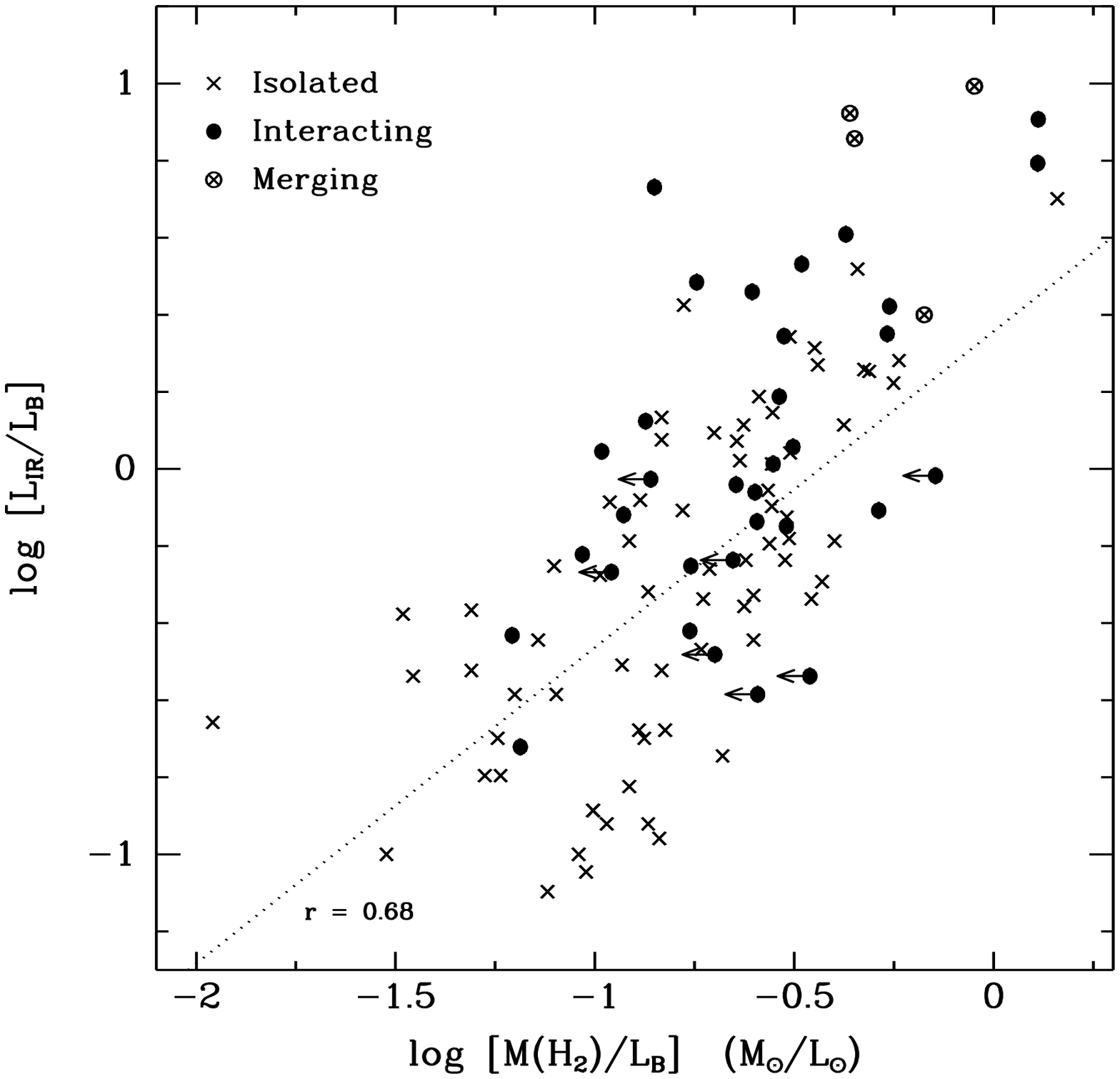,width=6.0in,angle=-90}
\end{center}
\caption{Far-infrared to blue luminosity ratio versus the molecular gas to
blue luminosity ratio for the isolated and interacting galaxies.
The dotted line is a linear fit to the isolated galaxy data only and $r$ is the
correlation coefficient.
All values for the interacting galaxies are averages for a system.
\label{fig_irb_h2b}}
\end{figure}

As was found in previous CO surveys, there is a very good correlation between
observed CO and far-IR fluxes and between derived \H2\ content and far-IR
luminosity for both samples of galaxies (Fig.~\ref{fig_ir_co}).
The fact that the correlation exists for the observed fluxes as well as the
derived luminosities and masses indicates that the
\LIR~versus~\MH2\ correlation is not simply a result of scaling the observed
fluxes by the square of the distance to each system.
It is also not simply due to far-IR emission and \H2\ content both increasing
with galaxy size because Fig.~\ref{fig_irb_h2b} shows that there is still a
strong correlation even after normalizing both \LIR\ and \MH2\ by \LB.
Many of the interacting systems---in particular all of the merging
systems---lie above the mean relations defined by the isolated galaxies
in Figs.~\ref{fig_ir_co} and \ref{fig_irb_h2b}.
Indeed the overall \LIR/\MH2\ and \LIR/\LB\ ratios for the interacting
systems are factors of $\sim$1.3 and $\sim$2.6 higher, respectively, than that
of the isolated galaxies.
The \LIR/\MH2\ and \LIR/\LB\ ratios also both tend to increase with increasing
fractional \H2\ content.

The correlation between fractional far-IR luminosity and \H2\ content
indicates that the current SFR in these galaxies is dependent upon the
molecular gas supply.
The fact that the interacting galaxies have a higher average \LIR/\MH2\ ratio
than the isolated galaxies further indicates that the amount of high-mass star
formation activity per unit molecular gas (or star formation efficiency, as it
is often called) is enhanced in these systems.

\begin{figure}
\begin{center}
\epsfig{file=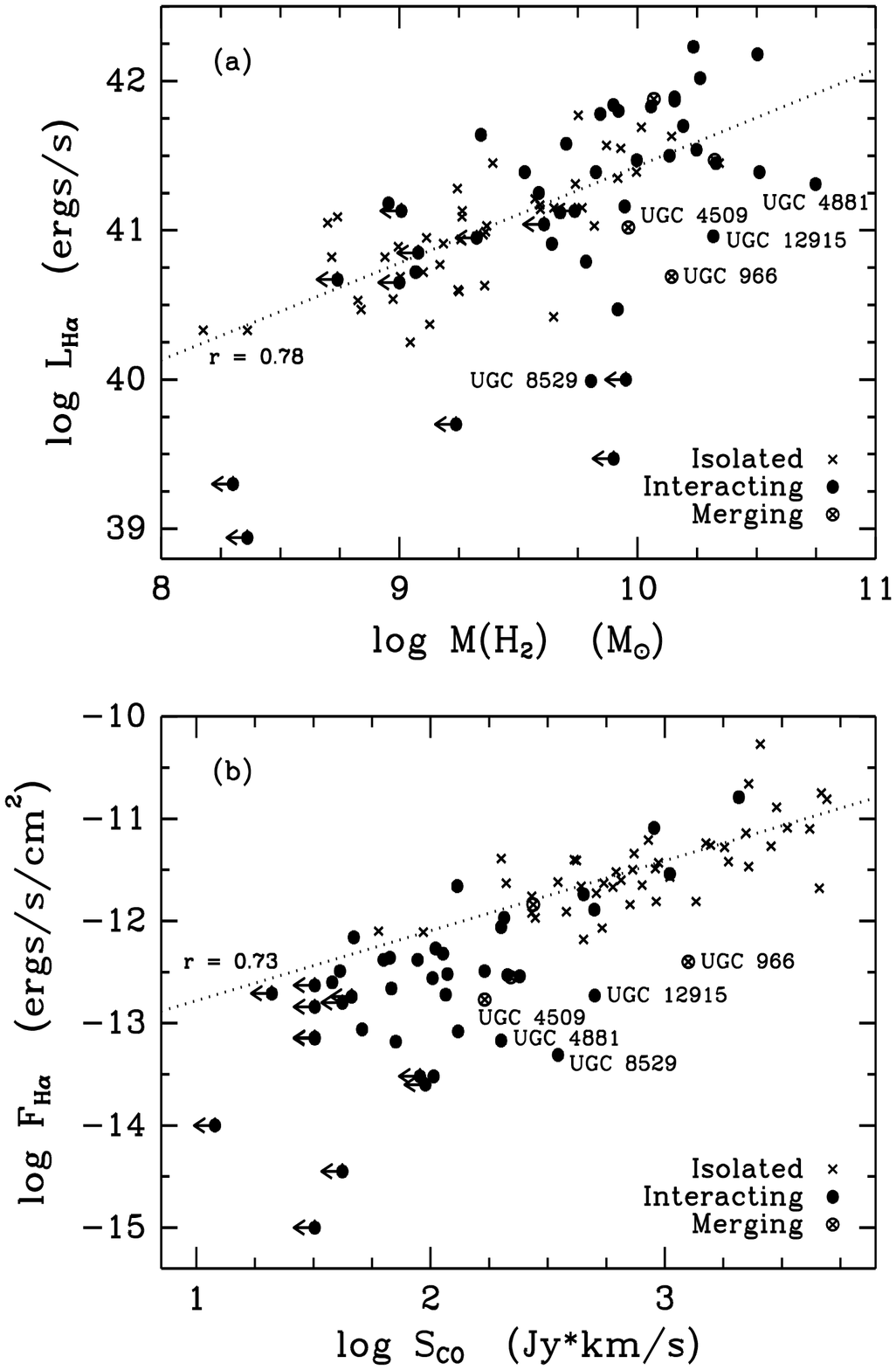,width=5.5in,angle=0}
\end{center}
\caption{(a) \Ha\ luminosity versus \H2\ content and (b) \Ha\ versus CO
emission fluxes for the isolated and interacting galaxies.
The dotted lines are linear fits to the isolated galaxy data only and $r$ is the
correlation coefficient.
Values for the interacting systems are for individual galaxies, when available,
otherwise they are totals for a system.
\label{fig_ha_co}}
\end{figure}

\begin{figure}
\begin{center}
\epsfig{file=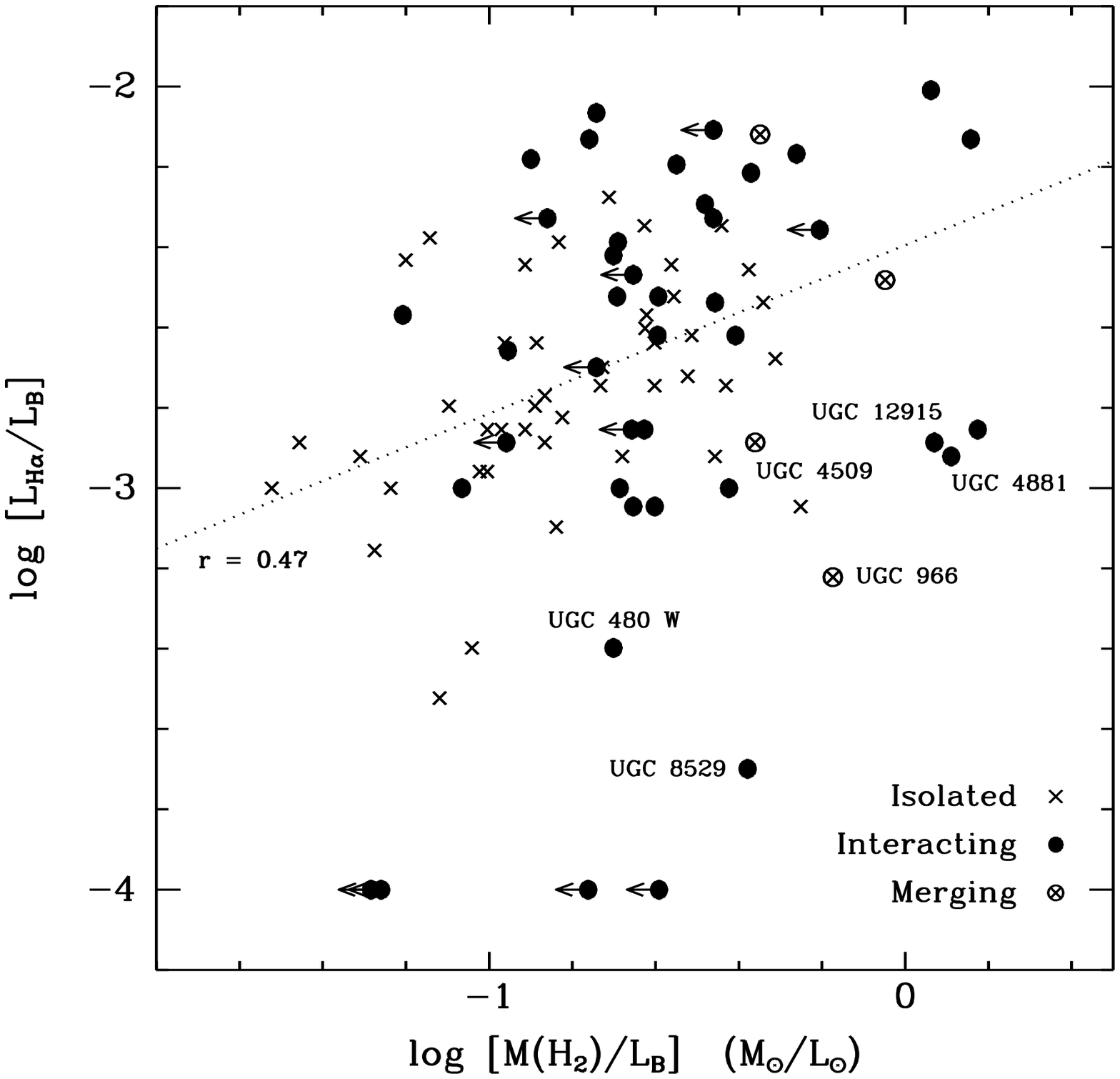,width=6.0in,angle=-90}
\end{center}
\caption{\Ha\ to blue luminosity ratio versus \H2\ gas mass to blue luminosity
ratio for the isolated and interacting galaxies.
The dotted line is a linear fit to the isolated galaxy data only and $r$ is the
correlation coefficient.
Values for the interacting systems are for individual galaxies, when available,
otherwise they are the global average for a system.
\label{fig_hab_h2b}}
\end{figure}

\Ha\ observations of the interacting systems have an advantage over the
\IRAS\ far-IR data in that \Ha\ measurements are available for the individual
galaxies within many of the pairs.
Comparing the molecular gas and star formation rates of the individual
galaxies is important because it has been shown that the two galaxies in an
interacting pair can often have very different SFR's (e.g. \cite{jos84};
\cite{bus87}; \cite{tel88}; \cite{xu91}).
Figure~\ref{fig_ha_co} shows the relationships between observed \Ha\ and CO
fluxes, as well as derived \Ha\ luminosities and \H2\ gas masses for the
samples of interacting and isolated galaxies.
In these figures the data for individual interacting galaxies have been plotted
when available, however the data for compact and merging systems are still
shown as global values.
These figures indicate that there is a good correlation between total
\Ha\ luminosity and derived \H2\ content for these galaxies.
A large part of this correlation, however, may be due to both \Ha\ emission
and \H2\ content scaling with galaxy size.
Normalization by \LB\ should remove much of this bias.
Figure~\ref{fig_hab_h2b} shows the relationship between \LHa/\LB\ and
\MH2/\LB\ for the two samples.
There is still a correlation, but a much weaker one (correlation coefficient
$r\sim 0.47$) than was seen when using \LIR\ as a measure of star formation.

\begin{figure}
\begin{center}
\epsfig{file=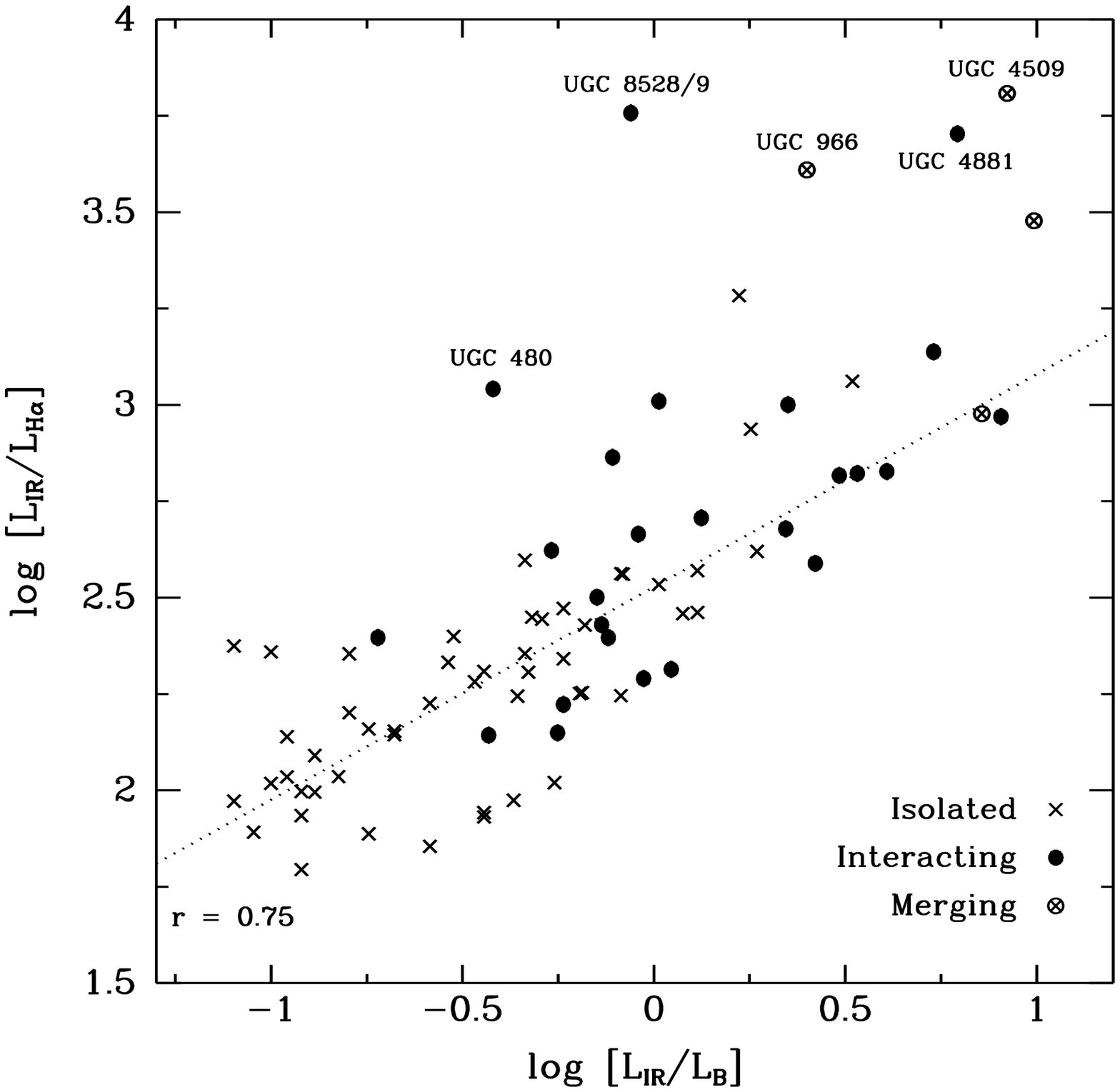,width=6.0in,angle=-90}
\end{center}
\caption{Far-infrared to \Ha\ luminosity ratio versus far-infrared to blue
luminosity ratio for the isolated and interacting galaxies.
The dotted line is a linear fit to the isolated galaxy data only and $r$ is the
correlation coefficient.
All values for the interacting systems are global averages for each system.
\label{fig_irha_irb}}
\end{figure}

The weaker correlation and increased scatter in the \LHa/\LB\ versus
\MH2/\LB\ relation is likely due to variable amounts of internal extinction
from galaxy to galaxy, which can have a dramatic effect on the
\Ha\ measurements.
Many of the interacting galaxies in particular fall well below the mean
relations between \Ha\ and \H2, having low levels of \Ha\ emission relative to
their \H2\ content.
A comparison of optical and near-infrared images of UGC~966 and UGC~12915, for
example, shows that these galaxies suffer large amounts of extinction at
optical wavelengths (\cite{bus90}).
Others systems, such as UGC~4509, UGC~4881, and UGC~8528/8529, have
unusually high \LIR/\LHa\ ratios, which suggests that much of the star
formation in these galaxies is occuring in embedded regions,
so that any \Ha\ emission associated with young stars is heavily extinguished.
The relationship between \LIR/\LHa\ and \LIR/\LB\ (Fig.~\ref{fig_irha_irb})
shows that most of these same galaxies scatter well above the mean relation,
confirming that they have low levels of \Ha\ emission relative to their far-IR
luminosities.
The high levels of far-IR emission from these galaxies indicates that they are
experiencing significant star formation, but the \Ha\ emission from the
young stars is largely being obscured by dust.

\subsubsection{Star Formation Rate Versus \H2\ Content for Individual Systems}

An examinination of the relationship between \H2\ content and star
formation activity for the two galaxies within individual interacting pairs
has also been performed.
As metioned above, the two galaxies within a given interacting pair often have
very different levels of current star formation activity, and so an analysis
of individual pairs will help to determine whether or not these disparities
are associated with differences in molecular gas content.
Far-IR and \Ha\ emission measurements are again used as star formation
indicators.

Unfortunately, few, if any, of the interacting pairs in this sample are resolved
in the \IRAS\ survey data, due to its low angular resolution.
Thus the preceeding comparisons of far-IR emission and molecular gas content
were based on global values for each system.
Bushouse, Telesco, \& Werner (1998), however, have resolved the far-IR emission
in six of the interacting pairs in the present sample using observations from
the Kuiper Airborne Observatory.
In addition, for some of the remaining widely-separated pairs, the \IRAS\ Point
Source Catalog (PSC) source location and error ellipse information can be used
as an indication of the relative distribution of far-IR emission in the pairs.
For example, in pairs where the PSC location and error ellipse are clearly
associated with and confined to one of the galaxies, we assume that this
galaxy is the dominant far-IR source.
In others, where the \IRAS\ PSC location is located between the optical
locations of the galaxies, we assume that the far-IR emission is distributed
more or less equally amongst the two galaxies.

There is a strong correlation between the relative levels of total far-IR and
CO emission for the two galaxies in each of these six systems observed by
Bushouse, Telesco, \& Werner (1998).
Four systems---UGC~5617/5620, UGC~7776/7777, UGC~8335, and UGC~11175---have
one galaxy that is the dominant far-IR source, and in all cases this galaxy is
also the dominant source of CO emission.
In the other two systems---NGC~4038/4039 and UGC~8641/8645---both far-IR and CO
emission are distributed approximately equally between the two galaxies
in the pair.
An analysis of the \IRAS\ PSC positions for the remaining pairs in the
samples shows that there are nine systems where the PSC position is clearly
associated with or dominated by one of the two galaxies in the pair: Arp~248,
Arp~256, UGC~480, UGC~6224, UGC~8528/8529, UGC~10923, UGC~11673,
UGC~12699/12700, and UGC~12914/12915.
In all of these pairs, the dominant far-IR galaxy is also the dominant or
only source of CO emission.
There are four pairs in which the far-IR emission is not clearly dominated by
either of the two galaxies: UGC~813/816, UGC~6471/6472, UGC~7938/7939, and
UGC~11284.
In two of these pairs---UGC~7938/7939 and UGC~11284---the observed CO emission
is also distributed nearly equally between the two galaxies.
In the UGC~813/816 and UGC~6471/6472 pairs, however, one of the two galaxies
is the dominant source of CO emission by factors of 1.7 and 2.1, respectively.

\Ha\ measurements are available for the individual galaxies within a total of
15 of the interacting pairs.
For all pairs in which one galaxy has a much higher current SFR---as indicated
by total \Ha\ emission---than its companion (e.g. Arp~248, UGC~480,
UGC~5617/5620, UGC~8335, UGC~10923, UGC~11175, and UGC~12699/12700) the active
galaxy has moderate to large amounts of molecular gas, while its inactive
companion has much lower or undetectable levels of \H2.
For pairs in which the two galaxies have similar SFR's (e.g. UGC~813/816,
UGC~7776/7777, UGC~7938/7939, UGC~8641/8645, UGC~11284, and UGC~12914/12915),
the two galaxies also have comparable levels of \H2\ content.

At first glance these data would seem to indicate a strong correlation between
molecular gas content and current star formation rates.
However, a large part of this correlation appears to be due to a simple scaling
of both parameters with overall galaxy size.
In nearly every one of the interacting pairs mentioned above as having one
galaxy as the dominant \Ha, far-IR, and CO source, this galaxy also has a
much larger optical blue luminosity than its companion.
In many cases the ratio of \LB\ values is comparable to the ratio of
\MH2\ values for the two galaxies.
In the UGC~480, UGC~5617/5620, UGC~10923, UGC~11175, and UGC~12699/12700 pairs,
for example, in which the \LB\ ratio of the two galaxies is a factor of 3 or
greater, it is always the more luminous galaxy that is the dominant \Ha,
far-IR, and CO emission source.
On the other hand, there are also pairs such as UGC~7776/7777 and UGC~8335,
in which the two galaxies have comparable optical luminosities and apparently
similar morphological types, yet one galaxy in the pair is clearly dominant in
\Ha, far-IR, and CO emission.

\begin{figure}
\begin{center}
\epsfig{file=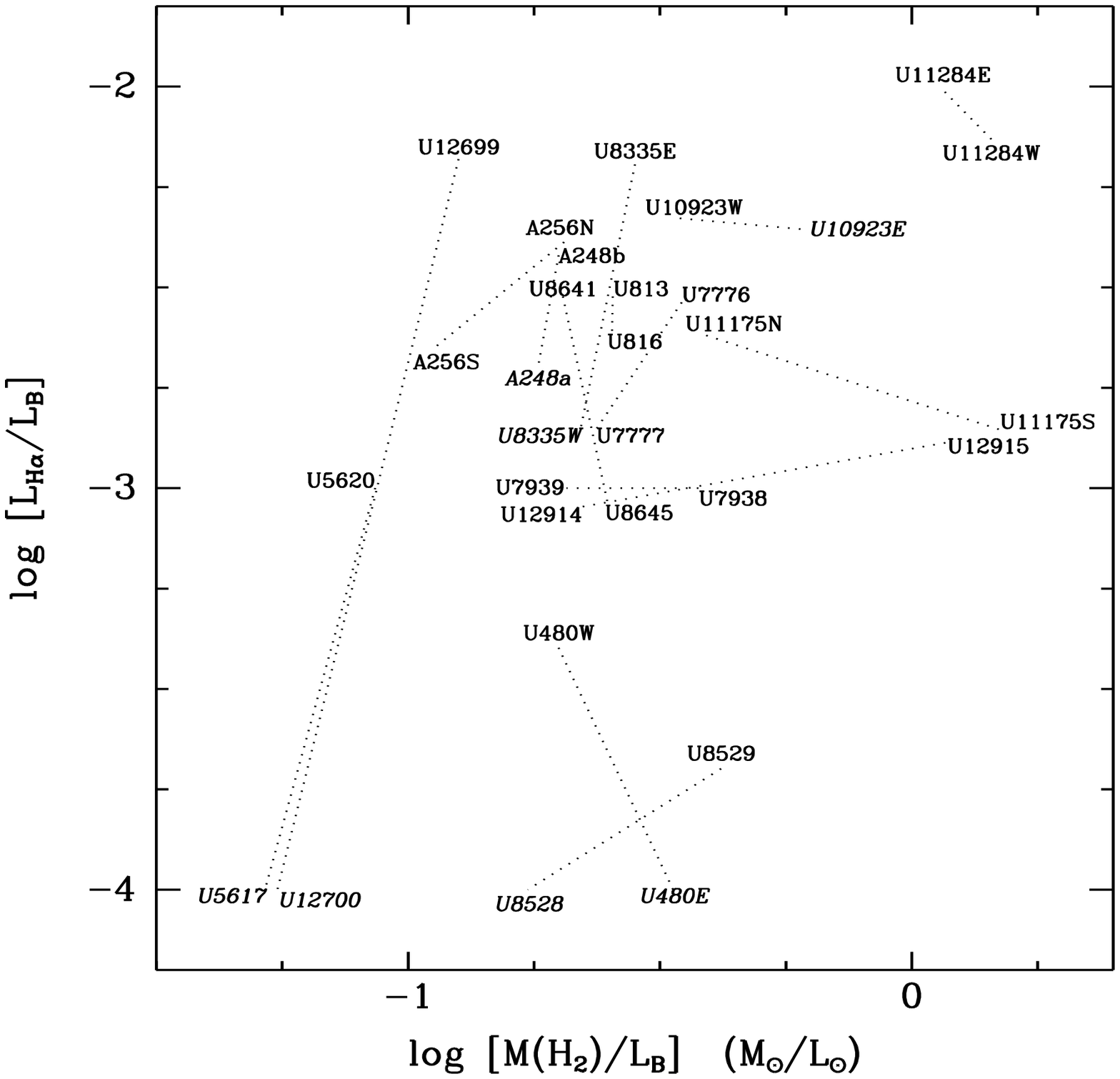,width=6.0in,angle=-90}
\end{center}
\caption{\Ha\ to blue luminosity ratio versus \H2\ gas mass to blue luminosity
ratio for individual interacting pairs.
All values are derived from measurements of the individual galaxies.
Object names in italic font indicate CO non-detections and therefore the
\MH2/\LB\ data point represents an upper-limit.
\label{fig_h2b_hab_names}}
\end{figure}

Therefore normalization by galaxy size is once again necessary in order
to reveal any real trends between molecular gas content and star formation
activity.
Using \LB\ again as a normalization parameter, Figure~\ref{fig_h2b_hab_names}
shows the \LHa/\LB\ ratios as a function of \MH2/\LB\ ratio for the 15
interacting pairs that have individual \Ha\ and CO measurements.
Seven of the pairs included in this figure have one galaxy with detectable CO
emission and the other without: Arp~248, UGC~480, UGC~5617/5620, UGC~8335,
UGC~8528/8529, UGC~10923, and UGC~12699/12700.
In each of these pairs the galaxy with detectable CO emission also has a
(often substantially) larger \LHa/\LB\ value.
Of the remaining eight pairs that have detectable CO emission in both
galaxies, only two---Arp~256 and UGC~7776/7777---show a strong correlation
between fractional \H2\ content and star formation, where the pair member
with the higher \MH2/\LB\ value also has a higher \LHa/\LB\ value.
In another three systems---UGC~813/816, UGC~7938/7939, and UGC~11284---the two
galaxies have comparable amounts of star formation activity and molecular gas.
The final three systems---UGC~8641/8645, UGC~11175, and UGC~12914/12915---do
not show a strong correlation between SFR and \H2\ content.
Thus a large fraction of the available sample shows a clear dependence between
current levels of star formation and molecular gas content, while there are
a few systems that do not.

\subsection{Subsamples Based on Star Formation Activity}

The sample of interacting galaxies consists of objects that span a large range
of star formation activity, as indicated by both their optical and infrared
properties.
In this section the mean properties of three subsamples of the
interacting galaxies are examined, where the subsamples are based on
star formation properties.
The subsamples are composed of galaxies with similar optical or infrared
properties, and hence an analysis of their mean properties will help to reduce
the inherent system-to-system scatter that has been seen in the preceding
analyses.

The three subsamples are taken from a study of the far-IR properties of
the parent sample of interacting galaxies by Bushouse, Lamb,
\& Werner (1988) and are listed in Table~\ref{tbl_subsmp}.
The first group, labeled ``High SFR-High IR'', are systems that have optical
indications of high current SFR's (e.g. \HII\ region like spectra and high
\Ha\ emission) and high levels of far-IR emission.
This group contains all of the merging systems in the interacting
sample from this study, as well as other very strongly interacting systems.
Systems in the ``High SFR-Low IR'' group, on the other hand, have optical
indications of high SFR's, but little or no far-IR emission detected by
\IRAS.
The ``Low SFR'' group consists of systems that have optical spectra with weak
or undetectable emissions lines and a stellar continuum dominated by older
G--M type stars.
Because the \IRAS\ data for the interacting systems do not resolve
the individual galaxies within the pairs, care has been taken to select
systems for these subsamples which contain galaxies with similar optical
properties.
The mean properties of the subsamples are listed in Table~\ref{tbl_subprop}.

\subsubsection{High SFR, High IR Emission Galaxies}

The galaxies in the High SFR-High IR group have all of the far-IR and
molecular gas properties typically seen in IR-selected samples of galaxies.
They have very high \LIR/\LB\ and \LIR/\MH2\ ratios, which are signs of
high SFR's and SFE's.
They also appear to be very rich in molecular gas, as evidenced by the factor
of four increase in the mean \MH2/\LB\ ratio as compared to the isolated
spirals.
The ratio of molecular to atomic gas is also enhanced by a factor of $\sim$3
in these systems.
The high \H2/\HI\ ratios are not due to lower levels of \HI\ content, as the
mean \MHI\ and \MHI/\LB\ are also both systematically higher for these systems
than for the isolated galaxies.

\subsubsection{High SFR, Low IR Emission Galaxies}

The High SFR-Low IR group galaxies, on the other hand, have very low levels of
far-IR emission and appear to be nearly devoid of molecular gas.
In fact none of these systems was detected in our CO observations.
In this respect they conform well to the far-IR versus CO emission
correlation.
The mean global \Ha\ luminosity---and hence the current population of young,
high-mass stars---in these systems is comparable to that of the isolated galaxy
sample.
The mean \LHa/\LB\ ratio, however, is equal to that of the High SFR-High IR
group of interacting galaxies and more than twice that of the isolated galaxies,
which indicates that a much larger fraction of the total stellar populations
of these galaxies is in young, high-mass stars.
This is confirmed by the very large \Ha\ emission-line equivalent widths seen
in the optical spectra of these galaxies (\cite{bus88}).
Yet they appear to have very little, if any, molecular gas to support this
star formation activity.
Thus the high-mass SFE, as measured by the \LHa/\MH2\ ratio, is the highest of
any of the groups.

The High SFR-Low IR systems also have low blue luminosities and therefore may
be intrinsically small or low-mass galaxies.
However, they do contain reasonably large amounts of atomic hydrogen,
resulting in a larger than normal mean \HI/\LB\ ratio.
The low blue luminosities of these galaxies suggests that they may be
low metallicity systems, since metallicity appears to be correlated with
galaxian mass (e.g. \cite{pag81}; \cite{gar87}; \cite{vil92}).
Emission line ratios in the optical spectra of these galaxies support this
conclusion.
The mean O/H ratio derived from emission line measurements is a factor of
3.5 smaller than that of the High SFR-High IR systems.
This is consistent with the lack of both far-IR emission and molecular gas.
Lower metallicity is expected to be accompanied by a smaller dust abundance,
which is necessary to reradiate the UV-optical light of high-mass stars
into the far-IR.
Since \H2\ is formed in a grain surface reaction, the lack of dust will
also lower the \H2\ formation rate.

Alternatively, recent theoretical and observational evidence suggests that
the CO/\H2\ ratio may be much lower than normal in low metallicity galaxies
(e.g. \cite{arn88}; \cite{mal88}; \cite{mad97}; \cite{smi97}).
Low abundances of C and O and a low dust column density may result in deeper
penetration of UV photons from high-mass stars, decreasing the size of the
CO-emitting region within a molecular cloud, while \H2\ remains relatively
self-shielded.
This can result in large volumes of molecular clouds which are not sampled
by CO observations, giving rise to an abnormally high CO-to-\H2\ conversion
factor.
If this is the case, then the derived \H2\ contents for the High SFR-Low IR
group of galaxies will be severly underestimated.

\subsubsection{Low SFR Galaxies}

The Low SFR subsample is comprised of galaxies with optical luminosities and
\HI\ gas masses comparable to, or greater than, most of those in the whole
isolated and interacting samples.
The current SFR's in these systems, however, are very low, as can be seen from
their low mean \Ha\ luminosities and \LHa/\LB\ ratios.
In fact the mean \LHa/\LB\ ratio is a factor of ten lower than that of the
isolated spirals.
Of the seven objects that have been observed in this subsample,
CO emission has been detected from only two galaxies (UGC~480~West and
UGC~11673~South).
Therefore it appears that one of the main reasons for the lack of current star
formation activity in these galaxies is a corresponding lack of molecular gas.
Notice however, that the two galaxies that were detected in CO, have 
considerable amounts of molecular gas---more than many of the very active
systems in the High SFR-High IR subsample.
Thus not all inactive systems can be explained by a simple lack of molecular
gas, but instead suggests that other specifics of the interaction events in
which they are involved have failed to provide the necessary inducement for
star formation activity.

\subsection{Subsamples Based on Interaction Strength}

Previous studies of the molecular gas properties of small samples of
interacting galaxies have indicated some correlations between interaction
strength and both \H2\ content and \LIR/\MH2\ ratios, where interaction
strength was based on the proximity of companions and the degree of
morphological disturbance.
For example, Solomon \& Sage (1988) found that weakly interacting galaxies
have average \LIR/\MH2\ and \MH2/\LB\ ratios that are not substantially
different than isolated galaxies, but strongly interacting and merging systems
have significantly higher values of both quantities.
Young et~al.~(1996) also found that the high-mass SFE, as measured by
\LHa/\MH2\ ratios, is enhanced by a significant amount in strongly 
interacting---and in particular merging---systems, while weakly interacting
pairs show no such increase.
These studies, however, were based upon either IR-selected samples of galaxies,
and thus may be biased towards galaxies with high molecular gas content, or
sampled a relatively small range of interaction types and small numbers of
systems.

\begin{figure}
\begin{center}
\epsfig{file=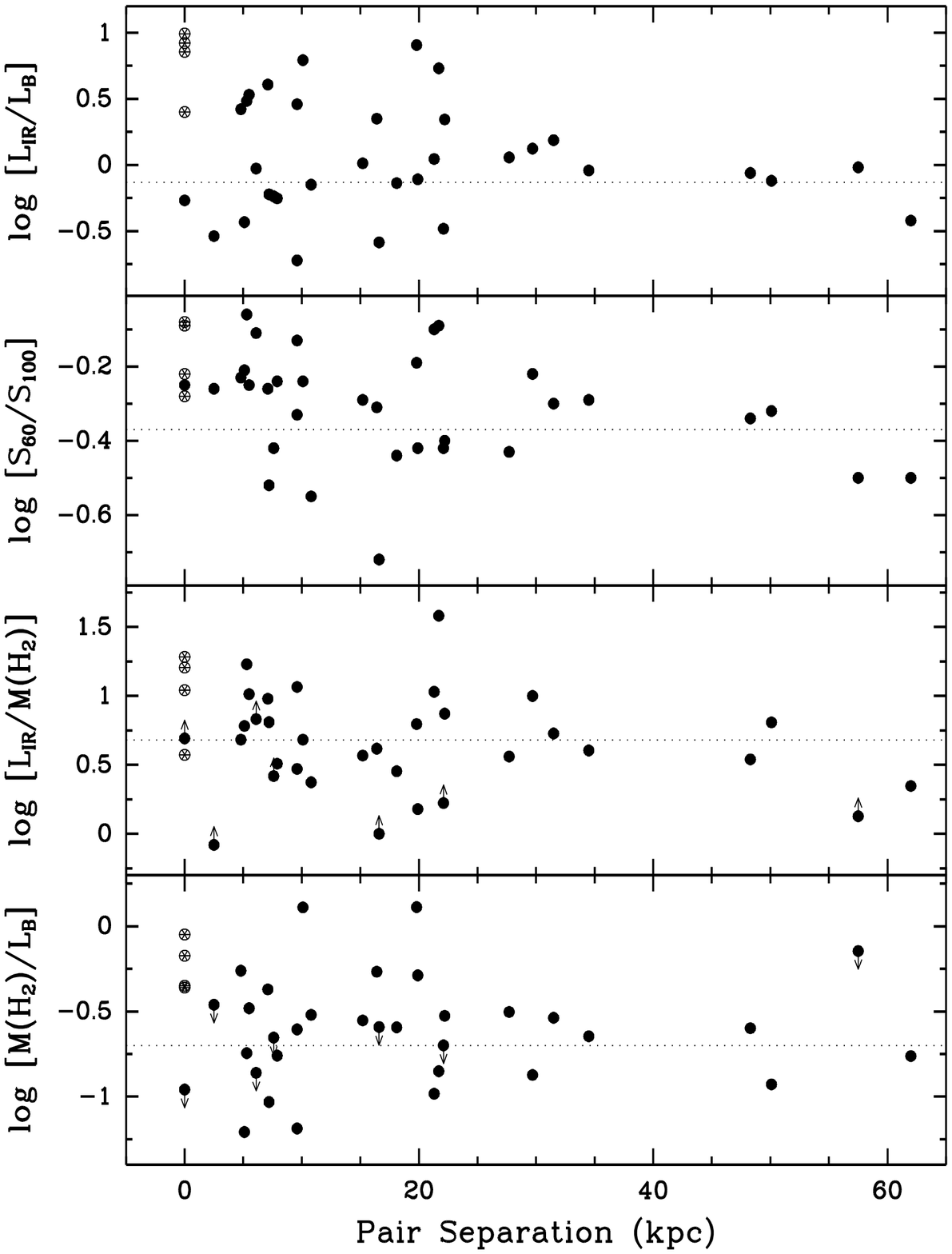,width=6.0in,angle=0}
\end{center}
\caption{Interacting galaxy characteristics as a function of projected pair
separation.
All values are global averages for each system.
The dotted line in each panel shows the mean value for the sample of isolated
galaxies.
\label{fig_all_sep}}
\end{figure}

We have performed a similar analysis using two different approaches.
First, the projected separation of the galaxies in each
interacting system has been used as a rough estimate of the strength or
severity of the collisions.
Figure~\ref{fig_all_sep} shows the relationship between pair separation and
several derived quantities for the interacting systems.
The \fsixty/\fhundred\ and \LIR/\LB\ ratios can both be used as indicators
of current SFR's due to the increase in dust temperature from heating
by massive stars, as well as a corresponding overall increase in far-IR
dust emission.
The relations shown in Figure~\ref{fig_all_sep} indicate that there is very
little correlation between SFR and pair separation.
There is only a modest increase in the average values of the
\fsixty/\fhundred\ and \LIR/\LB\ ratios as separation decreases.
The only connection that can be seen is the fact that no systems with large
separations have higher than normal SFR's.
Therefore small separations are a necessary but not sufficient condition for
significant interaction-induced star formation activity.

Figure~\ref{fig_all_sep} shows the same behavior for the \LIR/\MH2\ and
\MH2/\LB\ ratios as a function of pair separation.
There is only a modest increase in the average values as pair separation
decreases and no systems with large separations have elevated values.
The absence of any obvious correlation between pair separation and these
measures of star formation and molecular gas content indicates that either
there is no correlation between interaction strength and induced star formation
activity or that projected pair separation, by itself, is a poor indicator of
interaction strength.
The next approach will show that there is in fact a correlation between
interaction strength and star formation activity and therefore we conclude that
projected pair separation is simply a poor measure of interaction strength.
This is not surprising since the observed pair separation always provides only
a lower limit on the true separation due to projection effects.
Furthermore, we may be seeing some strongly interacting systems at a late
evolutionary stage when the two galaxies have already passed the point of
closest approach.

The second approach that has been used is to classify the strength of the
interactions based upon the morphological appearance of each system.
The systems have been classified into five groups, using the following criteria.
\begin{itemize}
\item{Class 1: }
Galaxies with nearby companions, but showing no obvious morphological
disturbance.
\item{Class 2: }
Galaxies with nearby companions and modest morphological disturbances, but
no tidal tails or bridges.
\item{Class 3: }
Galaxies with nearby companions and severe morphological disturbances,
including tidal tails, bridges, loops, and streamers.
\item{Class 4: }
Galaxies in apparent contact with their companions and with severe morphological
disturbances. Included in this category are a few systems which are not in
physical contact at this time, but obviously suffered a head-on or nearly
head-on collision in the recent past.
For example, the UGC~12699/12700 pair exhibits a ring feature which must have
been produced by a head-on collision, and is included in this category.
\item{Class 5: }
Merging systems, with a single amorphous body, possibly double nuclei, and a
pair of remnant tidal tails.
The four systems identified as mergers throughout the preceding analysis make
up this category.
\end{itemize}

\begin{figure}
\begin{center}
\epsfig{file=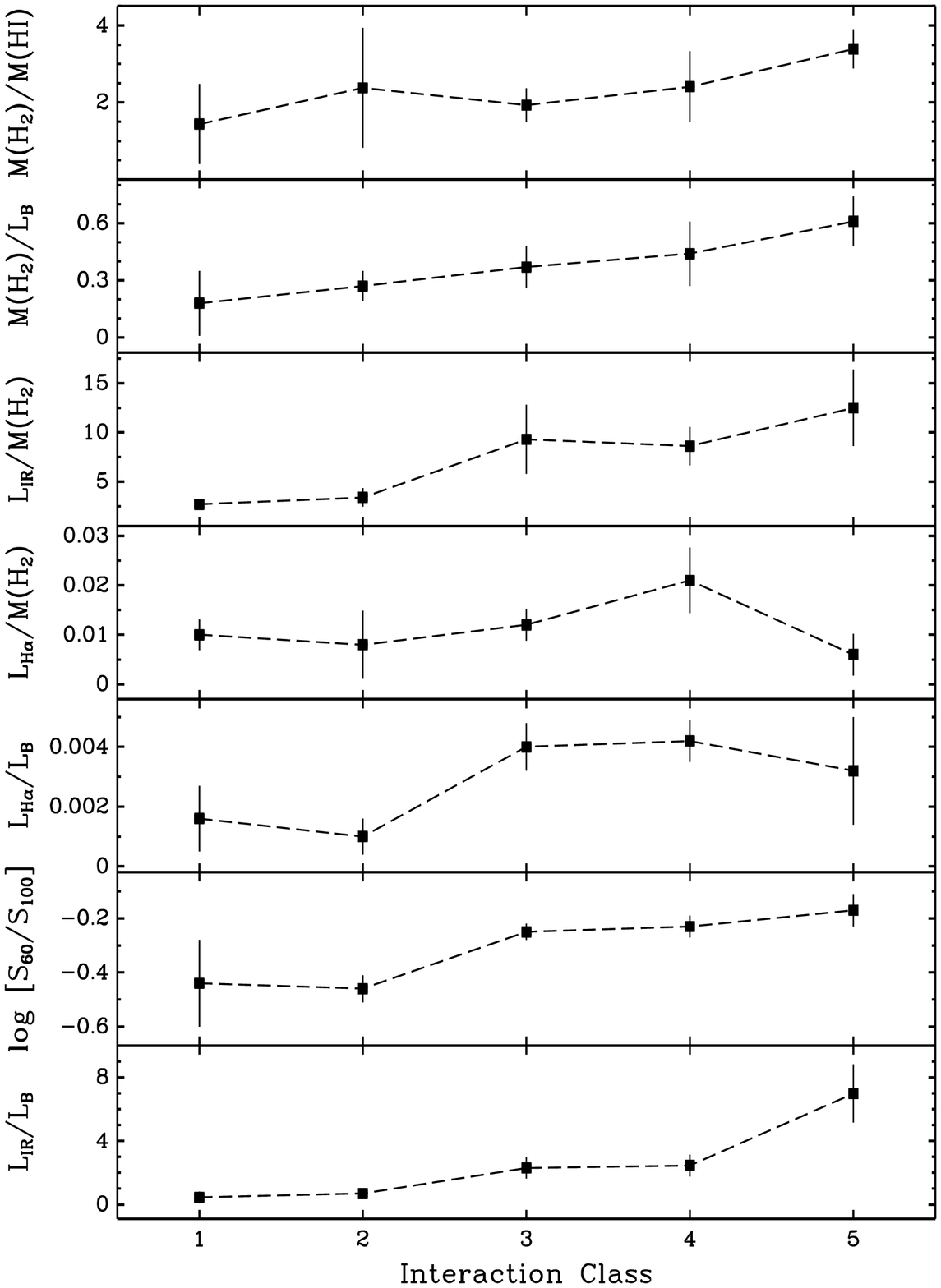,width=6.0in,angle=0}
\end{center}
\caption{Mean characteristics as a function of interaction class.
Error bars are the 1$\sigma$ uncertainty in the mean.
\label{fig_class_props}}
\end{figure}

Table~\ref{tbl_class} lists the interacting systems that have been assigned to
each class.
Table~\ref{tbl_clsprop} lists the mean properties of the classes, which are
also displayed graphically in Figure~\ref{fig_class_props}.
Several trends are evident.
First, the far-IR luminosity, \LIR/\LB\ ratio, and dust temperature
(as indicated by the \fsixty/\fhundred\ flux ratio) all increase systematically
with interaction strength.
The \LHa/\LB\ ratio shows a similar trend, but drops back down somewhat in
class 5 systems.
This is almost certainly due to extremely large extinction effects on the
observed \Ha\ emission in the very dusty class 5 merging systems.
All of these trends are a clear indication of increased mean SFR's
in the more strongly interacting systems.
Second, there is a steady increase in the \LIR/\MH2\ and \LHa/\MH2\ ratios
(except again for the \LHa/\MH2\ ratio in class 5 systems), which indicates
that the high-mass SFE also increases with interaction strength.
Finally, there is an obvious correlation between interaction class and both
\MH2/\LB\ and \MH2/\MHI\ ratios, indicating that the more strongly interacting
systems have systematically larger amounts of molecular gas per unit optical
luminosity and per unit atomic gas mass.
Some of the apparent increase in \MH2/\MHI\ for the class 5 systems could
be due to underestimates of their \HI\ emission fluxes.
All of these systems show an absorption component in their \HI\ spectra, due
to the presence of strong nuclear radio sources (\cite{bus87}; \cite{mir89}).

Notice that the largest changes in these properties often occur in the
transitions from class 2 to 3 and again from class 4 to 5.
The jump from class 2 to 3 indicates that a moderately strong interaction is
necessary to have any significant effect on the galaxies, and the transition
from class 4 to 5 indicates that mergers have the largest effect.
The more subtle changes from class 3 to 4 may be partially due to the highly
subjective nature of classifying the systems, which results in some mixing
between the classes.

It is also interesting to note that all of the interacting systems identified in
the previous section as having high SFR's---both those with and without
correspondingly high levels of far-IR emission---are members of class 3
or higher systems, with many being in classes 4 and 5.
The systems identified as having low current SFR's, however, are all in the
interaction strength class 2 category.
This adds further support to our conclusion that, in addition to having a
sufficient supply of molecular gas, it is necessary to have a moderately strong
interaction in order to stimulate high levels of star formation activity.

\section{Discussion}

Using the standard Galactic CO-to-\H2\ conversion factor to estimate
\H2\ content we have found that there is a good correlation between
fractional \H2\ content and current star formation rates for the overall
samples of isolated and interacting galaxies.
Furthermore, galaxies with high current SFR's almost always have moderate
to large amounts of \H2\ gas to support that activity; the only exceptions
being the low-metallicity galaxies with high SFR's but low far-infrared and
CO emission levels.
The fractional content of molecular gas for the entire sample of interacting
galaxies, as measured by both their \MH2/\LB\ and \MH2/\MHI\ ratios, is also
somewhat higher than that for the isolated galaxies, at a level of
$\sim$20--40\%.
The fractional \H2\ content of the interacting systems shows a steady increase,
on average, with increasing interaction strength, with the most strongly
interacting and merging systems having the largest increase relative to
isolated galaxies.

Previous studies have also examined the relationship between molecular gas
content and interaction strength or stage, which have yielded mixed results.
Young \& Knezek (1989) found enhanced molecular to atomic gas ratios in
close pairs and mergers when compared to isolated spirals.
In a study of infrared-luminous galaxies---most of which were in interacting
pairs---Mirabel \& Sanders (1989) also found that the \H2/\HI\ ratio increases
with interaction strength and is highest in merging systems.
On the other hand, more recent studies of infrared-luminous merging systems
have found that the CO luminosity (and hence derived \H2\ content)
is either not correlated with interaction stage (\cite{sol97}) or actually
decreases with decreasing pair separation (\cite{gao96}; \cite{gao99}).
Furthermore, in a study of the molecular gas properties of another
optically-selected sample of interacting galaxies \cite{hor99}\ find no
correlation between molecular gas fraction and degree of morphological
disturbance.

If their is an increase in \H2\ content in the more strongly interacting
systems, this would imply that \H2\ is somehow being created by the interaction
event, and that the process by which this occurs acts more strongly in the
more strongly interacting and late-stage systems.
Many models of interacting and merging systems have shown that a
combination of gravitational torques and dissipative collisions will remove
enough angular momentum from disk gas to produce radial inflows into the
central regions of galaxies (e.g. Noguchi 1988, 1991; \cite{ols90};
\cite{mih92}; \cite{bar92}, 1996; \cite{lam94}; \cite{ger96}; \cite{mih96}).
The increase in density of this centrally concentrated gas, as well as the
lower velocity dispersion that may result from settling into the deeper
potential well of the nucleus, may then give rise to enhanced
\HI-to-\H2\ conversion processes (\cite{youk89}; \cite{mir89}).
Some of the models also indicate that the largest gas inflows will occur
near the final stages of merging systems, consistent with our finding the
highest \H2\ contents in merging systems.
The fact that interferometric observations of CO emission have revealed that
the molecular gas in interacting galaxies is preferentially concentrated
towards their nuclei, and that it becomes increasingly concentrated as a
function of interaction strength (e.g. \cite{san88a}; \cite{sar91};
\cite{sco91}; \cite{lo97}; \cite{gao99}) also tends to support this scenario.

There is, however, no direct evidence yet for this suggested increase in the
\HI-to-\H2\ conversion process.
Furthermore, some models have shown that gaseous inflows are not necessarily
common to all interacting systems.
Simulations by Mihos \& Hernquist (1996), for example, show that galaxy
structure plays a role in regulating the gas inflow activity and that the
exact nature of an inflow depends on the progenitor bulge-to-disk ratio,
the evolutionary phase of the interaction, and the orbital geometry of the
encounter.
\cite{str97} and \cite{lam98} have shown that collisions that occur with a more
face-on geometry actually produce outwardly expanding gas flows, leaving a
central hole in the cool gas distribution of the disk of the target galaxy.
Even in the simulations that show inward gas flows, it is not well known how
much of this material will remain in the nuclear region, nor is there
sufficient resolution or knowledge of the detailed physics of this gas in the
existing simulations to know exactly what happens to this gas when it
reaches the nucleus.
There is also evidence that the spatial distributions of \HI\ and \H2\ gas
tend to become anti-correlated in interacting systems.
As mentioned above, observations have shown that the distribution of molecular
gas becomes increasingly concentrated towards the centers of galaxies as an
interaction progresses.
\HI gas, on the other hand, being loosely bound to the galactic disks,
tends to get thrown out to large radii (e.g. \cite{hib95}; \cite{wan98}),
leaving depleted levels of atomic gas in the central regions.
Therefore it may not be possible to explain the observed increases in
\H2\ content in the nuclear regions of interacting galaxies as the result
of enhanced \HI-to-\H2\ conversion, because there may not be sufficient
quantities of \HI\ near the nuclei to support that process.

All of the preceding results are based on the usual assumption that CO emission
serves as a tracer of \H2\ and that there is a connection between the
velocity-integrated brightness \ICO\ of the $^{12}$CO(1--0) line to the column
density N(\H2) of molecular hydrogen.
This includes the assumption that the CO--\H2\ connection can be expressed as
a constant ratio N(\H2)/\ICO\ in units of molecules cm$^{-2}$/(K~km~s$^{-1}$).
The fundamental calibration of this conversion ratio has been done using
various observations of Galactic molecular clouds in the solar neighborhood.
Recently, however, evidence has been accumulating that the CO--\H2\ conversion
factor may not be constant with position in a galaxy, nor from galaxy to galaxy.

For example, a radial gradient of more than a factor of 10 has been reported
in our Galaxy (\cite{sod95}), and the conversion factor for the central
regions of starburst galaxies---similar to many of the interacting galaxies
in our sample---may differ by as much as two orders of magnitude from that of
the dust clouds in the inner disks of quiescent spirals (e.g. \cite{cra85};
\cite{sta91}; \cite{sol97}; \cite{dow98}).
These studies suggest that the CO brightness in galaxy disks is largely an
excitation effect.
We see in general only the skins of the molecular clouds in the emission line
of $^{12}$CO(1--0).
The observed brightness will therefore depend directly on the excitation
temperatures and beam filling factors of the clouds.
Furthermore, because $\ICO/N(\H2) \propto T / \rho^{\frac{1}{2}}$, where $T$ is
the gas temperature and $\rho$ is the density (Dickman, Snell, \& Schloerb 1986;
\cite{mal88}; \cite{elm89}), an increase in the cloud temperatures will give
rise to higher CO luminosities per unit \H2\ mass, resulting in overestimates
of the \H2\ content.
Thus the $^{12}$CO(1--0) line will be bright where the ISM is warm, and not
necessarily where the \H2\ column densities are high.

Observations of nearby quiescent spiral galaxies indicate that the
molecular gas in these galaxies is likely to be generally cold and therefore
faint in CO emission (\cite{all97}; \cite{loi98}).
The skins of the molecular clouds may be sufficiently warmed by nearby UV
sources to be visible in the $^{12}$CO(1--0) line, but this emission will have
a small spatial filling factor, resulting in low observed intensities.
In galaxies with higher than normal levels of star formation, like many of the
interacting systems in our sample, the intense flux of UV photons from OB
stars together with cosmic-rays from supernovae will tend to dissociate the
low-density regions of molecular clouds, and the remaining high-density
regions will be heated, so that the CO emission lines will emanate from warm,
high-density clumps (\cite{all95}).
Studies of $^{12}$CO/$^{13}$CO and $^{12}$CO(2--1)/$^{12}$CO(1--0) flux ratios
in starburst and merging systems do in fact indicate that the CO emission is
coming from small, dense, and warm (T = 100--300 K) clouds 
(Aalto et~al.~1991a,b; \cite{aal95}).
The observed abundance ratios are significantly higher than those in giant
molecular clouds in the Galaxy, and also show radial variations within the
galaxies.

Finally, recent studies of the molecular ISM in ultraluminous infrared galaxies
have shown that the \MH2/\LCO\ ratio may be 3--5 times lower in the centers
of these galaxies than in Galactic molecular clouds (\cite{sol97};
\cite{dow98}).
Hence the application of the standard CO-to-\H2\ conversion factor leads to
significant overestimates of the amount of \H2\ in these galaxies.

It is possible, therefore, that at least some portion of the effects
seen in this study are due to changes in the properties of the molecular gas
within the sample galaxies rather than changes in the gas content itself.
Instead of the level of star formation activity being driven by molecular
gas content, the deduced molecular gas content may be at least partially
driven by the level of star formation activity.
Portions of the sample of isolated spirals and the subsamples of weakly
interacting systems have systematically low levels of current star formation
activity.
Thus the molecular gas in these galaxies will be cold and the observed level
of CO emission low, with potentially large amounts of \H2\ remaining unseen.
Meanwhile the more strongly interacting systems have overall higher current
star formation rates, which may lead to higher observed levels of CO emission
due to increased cloud heating provided by the larger population of young,
massive stars.

The increase in observed CO emission in the more active galaxies may be a
simple result of having more sites of active star formation within the galaxies,
thus increasing the filling factor of warm, CO-emitting regions.
At the same time, the large UV and cosmic-ray fluxes provided by massive stars
will tend to dissociate the \H2\ regions of the clouds, leading to an
abnormally high \ICO/N(\H2) ratio and a corresponding overestimate of the
mass of \H2\ gas.
The data in Tables~\ref{tbl_subprop} and \ref{tbl_clsprop} do in fact show that
the average dust temperatures increase with both interaction strength and
level of star formation activity.
The temperature increase, however, is not very large and the precise connection
between observed dust temperatures and molecular gas cloud excitation
temperatures is not well known.

The actively star-forming interacting galaxies in the sample studied here
appear to have properties that are intermediate between those of the
isolated galaxies and the ultraluminous infrared galaxies in other studies.
Thus we assume that an intermediate CO-to-\H2\ conversion factor would be
appropriate for these systems.
Conversion factors 3 to 5 times lower than the standard have been suggested
for the ultraluminous infrared systems, therefore a conversion factor
$\sim$2 times lower than the standard may be appropriate for the active
systems in this study.

A reduction of a factor of 2 in the derived \H2\ masses for the galaxies
with high SFR's would easily account for the increases in mean \MH2/\LB\ and
\MH2/\MHI\ ratios between the low and high SFR subsamples of the interacting
systems.
Similarly, a factor of 2 reduction in \MH2\ for the active interacting class 5
systems would also account for most of the increase in \MH2/\LB\ and
\MH2/\MHI\ ratios from class 1 to class 5 systems.

However, changes in the conversion factor of this order cannot account for the
whole range of fractional \H2\ content seen within the entire isolated and
interacting galaxy samples.
The \MH2/\LB\ and \MH2/\MHI\ ratios for individual systems cover a range of a
factor of 100 in both quantities.
Therefore a change in the CO-to-\H2\ conversion factor on the order of 2
over the range of the samples cannot account for the strong correlation seen
between fractional \H2\ content and SFR, as measured by \LIR/\LB\ and
\LHa/\LB\ ratios (see, e.g., Figs.~\ref{fig_irb_h2b} and~\ref{fig_hab_h2b}).
Therefore even if the apparent increase in the \H2\ content of the interacting
galaxies---for the sample as a whole relative to the isolated spirals and as a
function of SFR and interaction strength---is not real, there still is a
correlation between the \H2\ content and the current SFR in individual
galaxies, for both the isolated and interacting systems.

\section{Summary}

We have analyzed CO (1--0) emission-line observations for a sample of 37
interacting galaxy systems.
The parent sample of interacting systems was compiled on the basis of
morphological evidence for current involvement in an interaction, without
bias towards known optical, infrared, or radio flux levels or indicators
of high current SFR's or nuclear activity.
The systems in this study were selected from the parent sample in
order to span a large range of interaction strengths and star formation
properties.
If we assume that the CO-to-\H2\ conversion factor is approximately constant
from galaxy to galaxy within our samples of interacting and isolated galaxies,
our results indicate the following.
\begin{itemize}
\item{1)}
There is a good correlation between fractional \H2\ content and current
SFR's for the overall samples of isolated and interacting galaxies, which
indicates that the level of star formation activity is dependent upon the
available molecular gas supply in these galaxies.

\item{2)}
Galaxies with high SFR's almost always have moderate to large amounts of
\H2\ gas and galaxies with low SFR's usually have little to no detectable \H2.
The only exceptions to this are the small number of systems in our
``High SFR-Low IR'' subsample of interacting systems, which have high current
SFR's but little or no detectable far-IR emission or \H2\ gas, and a few
members of the ``Low SFR'' subsample which appear to have moderate amounts of
\H2, but very low current SFR's.
The apparent lack of \H2\ gas in the ``High SFR-Low IR'' systems may be
explained by an anomalous \H2/CO ratio for the low metallicity ISM in these
galaxies.
Furthermore, the low metallicity ISM may also be deficient in dust, explaining
the lack of thermal far-IR emission from these galaxies.
A moderate to large molecular gas supply is therefore a necessary, but not
sufficient, condition for spawning significant levels of interaction-induced
star formation activity in galaxies.

\item{3)}
A comparison of the relative levels of star formation and \H2\ content in the
two galaxies within individual interacting pairs shows a correlation
between these two properties, but it is weaker than the correlation seen for
the overall samples.
This indicates that the lack of star formation activity sometimes seen in the
individual galaxies of interacting pairs is mainly due to a lack of molecular
gas, but also that the specific properties of each interaction event and
the galaxies involved must also play a role in governing the level of
interaction-induced star formation.
This is consistent with some numerical simulations of interacting galaxies
that show that interaction-induced gas flows within the galaxies are governed
by properties such as bulge-to-disk ratios, collision geometries, and the mass
ratios of the galaxies.

\item{4)}
The entire sample of interacting galaxies has a \LIR/\MH2\ ratio that is a
factor of $\sim$1.3 higher than that of the isolated spirals, indicating that
the rate of high-mass star formation per unit molecular gas---or star
formation efficiency---is enhanced in these systems.

\item{5)}
The fractional \H2\ content of the whole sample of interacting galaxies, as
measured by \MH2/\LB\ and \MH2/\MHI\ ratios, is $\sim$20--40\% higher, on
average, than that of the comparison sample of isolated spirals.
There is a strong correlation between fractional \H2\ content and interaction
strength, as determined by degree of morphological disturbance and companion
proximity.
There is a factor of 2--3 increase in \MH2/\LB\ and \MH2/\MHI\ ratios from
the most weakly interacting to the most strongly interacting systems in our
sample.
This implies that \H2\ is somehow being created in interaction events, and
that the process by which this occurs is most pronounced in the most strongly
interacting systems.
It has been suggested that this is a result of inflows of atomic gas to the
central regions of interacting galaxies, and that the large pool of
centrally-concentrated dense, cool gas may then be easily converted from
atomic to molecular form.
\end{itemize}

There are indications, however, that the CO-to-\H2\ conversion factor may
not be constant either within individual galaxies or from galaxy to galaxy.
Furthermore, the level of observed CO emission itself may be dependent on
the level of star formation activity within a galaxy, which will tend to
increase the ``visibility'' of molecular clouds.
Reductions in the CO-to-\H2\ conversion factor that have been suggested for
some ultraluminous infrared galaxies would be more than enough to nullify
the average increases in derived \H2\ content for the various subsamples of
interacting galaxies in this study.
They cannot, however, account for the total range of \H2\ content seen within
the entire samples of isolated and interacting galaxies.
Therefore the observed correlation between molecular gas content and current
star formation rates must be real.

\acknowledgments
S.A.L. wishes to thank the Aspen Center for Physics for hospitality during
the completion of this work.
Portions of this work were carried out at the Jet Propulsion Laboratory,
California Institute of Technology, under contract with the National
Aeronautics and Space Administration.
This research has made use of the NASA/IPAC Extragalactic Database (NED), which
is operated by the Jet Propulsion Laboratory, California Institute of
Technology, under contract with the National Aeronautics and Space
Administration.

\clearpage

\begin{deluxetable}{llcccccc}
%\tablenum{1}
\tablecolumns{8}
\tablecaption{CO data for interacting systems. \label{tbl_codata}}
\tablewidth{0pt}
\tablehead{
\colhead{\ }     & \colhead{Alternate} & \colhead{\ }   &\colhead{\VCO} &
\colhead{$\Delta$V$_0$} & \colhead{$\Delta$V$_{50}$} & \colhead{\SCO}  &
\colhead{\ } \\
\colhead{Object} & \colhead{Name}      & \colhead{Type} &\colhead{(\kms)}&
\colhead{(\kms)} & \colhead{(\kms)} & \colhead{(Jy \kms)} & \colhead{Ref} \\
\colhead{(1)} & \colhead{(2)} & \colhead{(3)} & \colhead{(4)} & \colhead{(5)} &
\colhead{(6)} & \colhead{(7)} & \colhead{(8)} }
\startdata
Arp 248 a   &\nodata  &complete&  5270    & \nodd & \nodd & \lt42\phd &1\nl
Arp 248 b   &\nodata  &pair    &  5120    &   370 &   160 &\phn67     &1\nl
\tablevspace{0.1in}
Arp 250 W   &\nodata  &1 of 2  & 24200\phn& \nodd & \nodd & \lt32\phd &1\nl
\tablevspace{0.1in}
Arp 256 N   &\nodata  &complete&  8150    &   160 &   100 &\phn38     &1\nl
Arp 256 S   &\nodata  &pair    &  8150    &   330 &   270 &\phn63     &1\nl
\tablevspace{0.1in}
NGC 1614    &Arp 186  &merger  &  4770    &   390 &   200 &   274     &1\nl
\tablevspace{0.1in}
NGC 4038    &\nodata  &complete&  1540    & \nodd &   180 &  1150\phn &2\nl
NGC 4039    &\nodata  &pair    &  1570    & \nodd &   230 &   920     &2\nl
\tablevspace{0.1in}
NGC 7592    &\nodata  &compact &  7340    &   440 &   300 &   105     &1\nl
\tablevspace{0.1in}
UGC 480 W   &\nodata  &complete& 11060\phn&   680 &   330 &   131     &1\nl
UGC 480 E   &\nodata  &pair    & 11130\phn& \nodd & \nodd & \lt32\phd &1\nl
\tablevspace{0.1in}
UGC 813     &\nodata  &complete&  5150    &\ap590\phs&\ap390\phs&\ap68\phn&1\nl
UGC 816     &\nodata  &pair    &  5280    &   490 &   270 &   118     &1\nl
\tablevspace{0.1in}
UGC 966     &NGC 520  &merger  &  2310    & \nodd &   270 &  1260\phn &2\nl
\tablevspace{0.1in}
UGC 993     &\nodata  &compact &  2880    & \nodd & \nodd & \lt32\phd &1\nl
\tablevspace{0.1in}
UGC 1720    &IC 214   &compact &  9080    &   520 &   210 &\phn88     &1\nl
\tablevspace{0.1in}
UGC 2320 SW &Arp 190  &1 of 2  & 10390\phn& \nodd & \nodd & \lt42\phd &1\nl
\tablevspace{0.1in}
UGC 2992    &\nodata  &compact &  5030    & \nodd & \nodd & \lt21\phd &1\nl
\tablevspace{0.1in}
UGC 3031    &NGC 1568a&1 of 2  &  4690    & \nodd & \nodd & \lt39\phd &1\nl
\tablevspace{0.1in}
UGC 3737    &\nodata  &compact &  4440    & \nodd & \nodd & \lt32\phd &1\nl
\tablevspace{0.1in}
UGC 4509    &NGC 2623 &merger  &  5310    & \nodd &   170 &   170     &2\nl
\tablevspace{0.1in}
UGC 4718 N  &NGC 2719 &1 of 2  &  3160    &   220 & \nodd &\phn47     &3\nl
\tablevspace{0.1in}
UGC 4744 W  &NGC 2735 &1 of 2  &  2450    &   230 & \nodd &\phn35     &3\nl
\tablevspace{0.1in}
UGC 4757    &NGC 2744 &compact &  3440    &\ap740\phs&\ap290\phs&\ap46\phn&1\nl
\tablevspace{0.1in}
UGC 4881    &Arp 55   &compact & 11950\phn& \nodd &   250 &   200     &2\nl
\tablevspace{0.1in}
UGC 5617    &NGC 3226 &complete&  1360    & \nodd & \nodd & \lt90\phd &2\nl
UGC 5620    &NGC 3227 &pair    &  1110    & \nodd &   340 &   450     &4\nl
\tablevspace{0.1in}
UGC 6224 N  &NGC 3561a&complete&  8740    &   450 &   320 &   125     &1\nl
UGC 6224 S  &NGC 3561b&pair    &  8730    & \nodd & \nodd & \lt49\phd &1\nl
\tablevspace{0.1in}
UGC 6471    &IC 694   &complete&  3160    & \nodd &   250 &   610     &2\nl
UGC 6472    &NGC 3690 &pair    &  2930    & \nodd &   260 &   290     &2\nl
\tablevspace{0.1in}
UGC 7776    &NGC 4568 &complete&  2220    &   260 & \nodd &  1050\phn &2\nl
UGC 7777    &NGC 4567 &pair    &  2280    &   180 & \nodd &   500     &2\nl
\tablevspace{0.1in}
UGC 7938    &NGC 4676a&complete&  6640    & \nodd &   390 &\phn71     &5\nl
UGC 7939    &NGC 4676b&pair    &  6570    & \nodd &   310 &\phn51     &5\nl
\tablebreak
%\tablevspace{0.1in}
UGC 8135    &NGC 4922 &compact &  7060    &   360 &   200 &\phn74     &1\nl
\tablevspace{0.1in}
UGC 8335 W  &Arp 238 W&complete&  9110    & \nodd & \nodd & \lt32\phd &1\nl
UGC 8335 E  &Arp 238 E&pair    &  9310    &   280 &   220 & \ap41     &1\nl
\tablevspace{0.1in}
UGC 8387    &IC 883   &merger  &  6960    & \nodd &   370 &   220     &2\nl
\tablevspace{0.1in}
UGC 8528    &NGC 5216 &complete&  2910    & \nodd & \nodd & \lt95\phd &1\nl
UGC 8529    &NGC 5218 &pair    &  2940    &   510 &   220 &   350     &1\nl
\tablevspace{0.1in}
UGC 8641    &NGC 5257 &complete&  6810    &   530 &   340 &   206     &1\nl
UGC 8645    &NGC 5258 &pair    &  6780    &   530 &   390 &   240     &1\nl
\tablevspace{0.1in}
UGC 10267   &NGC 6090 &compact &  8870    & \nodd &   120 &   200     &2\nl
\tablevspace{0.1in}
UGC 10923 W &\nodata  &complete&  7920    &   220 &\phn80 &   113     &1\nl
UGC 10923 E &\nodata  &pair    &  6630    & \nodd & \nodd & \lt32\phd &1\nl
\tablevspace{0.1in}
UGC 11175 N &NGC 6621 &complete&  6170    &   570 &   290 &   170     &1\nl
UGC 11175 S &NGC 6622 &pair    &  6230    &   470 &   300 &   103     &1\nl
\tablevspace{0.1in}
UGC 11284 W &NGC 6670a&complete&  8550    &   570 &   400 &   116     &1\nl
UGC 11284 E &NGC 6670b&pair    &  8690    &   370 &   230 &   102     &1\nl
\tablevspace{0.1in}
UGC 11673 N &\nodata  &complete& 14270\phn& \nodd & \nodd & \lt32\phd &1\nl
UGC 11673 S &\nodata  &pair    & 14300\phn&   520 &   370 &\phn42     &1\nl
\tablevspace{0.1in}
UGC 12699   &NGC 7714 &complete&  2860    & \nodd &   100 &   130     &2\nl
UGC 12700   &NGC 7715 &pair    &  2770    & \nodd & \nodd & \lt12\phd &3\nl
\tablevspace{0.1in}
UGC 12914   &\nodata  &complete&  4390    &   690 &   530 &   213     &1\nl
UGC 12915   &\nodata  &pair    &  4490    &   620 &   320 &   502     &1\nl
\enddata
\tablecomments{Col.~(1) Galaxy name as used in this paper.
Col.~(2) Alternate name from NGC, Arp, or IC catalogs.
Col.~(3) Type of system: ``complete pair'' - CO data for each of two
galaxies; ``1 of 2'' - CO data for one of two galaxies; ``compact'' -
single CO observation of entire system; ``merger'' - single CO observation of
advanced merger.
Col.~(4) The intensity-weighted mean velocity of the CO line, in units of
\kms. Entries for non-detections are the velocity center of the observation.
Col.~(5) The full width of the CO line, measured at the zero flux level.
Col.~(6) The full width of the CO line, measured at 50\%\ of the peak line
temperature.
Col.~(7) The CO integrated flux, in units of Jy \kms. Upper limits are
3$\sigma$.
Col.~(8) Sources of the CO data: (1) this paper, (2) Young et~al.~1995,
(3) Sofue et~al.~1993, (4) Bieging et~al.~1981, (5) Casoli et~al.~1996.
}
\end{deluxetable}

\clearpage

\begin{table}
\dummytable\label{tbl_obsdata}
\end{table}

\begin{table}
\dummytable\label{tbl_derdata}
\end{table}

\begin{deluxetable}{lccccc}
%\tablenum{4}
\tablecolumns{6}
\tablecaption{Mean and median sample properties.\label{tbl_mnprop}}
\tablewidth{0pt}
\tablehead{
\colhead{} & \multicolumn{2}{c}{Isolated Galaxies} & \colhead{} &
 \multicolumn{2}{c}{Interacting Galaxies\tablenotemark{a}} \\
 \cline{2-3} \cline{5-6} \\
\colhead{Quantity} & \colhead{Mean\tablenotemark{b}} & \colhead{Median} & \colhead{} &
 \colhead{Mean\tablenotemark{b}} & \colhead{Median} }
\startdata
D\ (Mpc)                & 23.0\mp2.3      & 15.0  & & 85.6\mp6.9     & 73.7 \nl
\LB\ (10$^9$ \Lsun)     & 19.5\mp2.2      & 12.6  & & 28.9\mp3.4     & 23.6 \nl
\LIR\ (10$^9$ \Lsun)    & 19.2\mp4.3      &  4.1  & & 87.1\mp14.0    & 50.7 \nl
\MHI\ (10$^9$ \Msun)    & 3.5\mp0.5       &  2.0  & & 7.4\mp1.0      &  5.2 \nl
\MH2\ (10$^9$ \Msun)    & 4.7\mp0.8       &  1.8  & & 9.9\mp1.4      &  6.4 \nl
log \LHa\ (ergs/s)      & 41.12\mp0.06    & 40.95 & & 41.57\mp0.10   & 41.18 \nl
log \fsixty/\fhundred   &$-$0.37\mp0.02   &$-$0.36& &$-$0.30\mp0.02  &$-$0.29\nl
\Tdust\ (K)             & 33.6\mp0.4      & 33.4  & & 35.7\mp0.7     & 35.5 \nl
\LIR/\LB                & 0.74\mp0.09     & 0.46  & & 2.30\mp0.42    & 1.00 \nl
\LHa/\LB                & 0.0022\mp0.0002 & 0.0018& & 0.0033\mp0.0004& 0.0023\nl
\LIR/\MH2\ (\Lsun/\Msun)& 4.8\mp0.5       &  3.4  & & 6.8\mp1.1      &  4.1 \nl
\LHa/\MH2\ (\Lsun/\Msun)& 0.016\mp0.002   & 0.011 & & 0.016\mp0.002  & 0.012 \nl
\MHI/\LB\ (\Msun/\Lsun) & 0.24\mp0.03     & 0.15  & & 0.29\mp0.04    & 0.22 \nl
\MH2/\LB\ (\Msun/\Lsun) & 0.20\mp0.02     & 0.15  & & 0.36\mp0.05    & 0.23 \nl
\MH2/\MHI               & 1.98\mp0.35     & 1.01  & & 1.85\mp0.26    & 1.45 \nl
\enddata
\tablenotetext{a}{Quantities involving \IRAS\ measurements are global values
for each interacting system.
All others use values for individual galaxies, when available.}
\tablenotetext{b}{Mean values have been computed using the Kaplan-Meier
estimator of a randomly censored distribution, which properly accounts for
non-detections. The quoted errors are the 1$\sigma$ uncertainty in the mean.}
\end{deluxetable}

\begin{deluxetable}{lll}
%\tablenum{5}
\tablecolumns{3}
\tablecaption{Interacting galaxy subsample members. \label{tbl_subsmp}}
\tablewidth{0pt}
\tablehead{
\colhead{High SFR-High IR} & \colhead{High SFR-Low IR} &\colhead{Low SFR} }
\startdata
NGC~1614    & UGC~993  & Arp~250~W   \nl
UGC~4509    & UGC~2992 & UGC~480~W   \nl
UGC~4881    & UGC~3737 & UGC~480~E   \nl
UGC~6471/2  & UGC~4757 & UGC~2320~SW \nl
UGC~8335    &          & UGC~3031    \nl
UGC~8387    &          & UGC~11673~N \nl
UGC~11284   &          & UGC~11673~S \nl
\enddata
\end{deluxetable}

\setcounter{page}{20}

\begin{deluxetable}{lccc}
%\tablenum{6}
\tablecolumns{4}
\tablecaption{Mean properties of interacting subsamples selected by SFR.\label{tbl_subprop}}
\tablewidth{0pt}
\tablehead{
\colhead{\ } & \colhead{\ } & \colhead{High SFR,} &\colhead{High SFR,} \\
\colhead{Quantity\tablenotemark{a}} & \colhead{Low SFR} & \colhead{Low IR} &\colhead{High IR} }
\startdata
Number of systems        & 5                & 4                & 7               \nl
\LB\ (10$^9$ \Lsun)      & 48.0\mp21.6      & 5.9\mp1.9        & 40.7\mp10.8     \nl
\LIR\ (10$^9$ \Lsun)     & 39.0\mp15.7      & \lt3.9\mp1.6     & 232.8\mp18.1    \nl
\MHI\ (10$^9$ \Msun)     & 6.6\mp2.9        & 3.8\mp0.7        & 9.7\mp4.5       \nl
\MH2\ (10$^9$ \Msun)     & \lt16.8\mp5.3    & \lt0.9\mp0.2     & 22.2\mp7.1      \nl
log \LHa\ (ergs/s)       & 40.83\mp0.30     & 40.88\mp0.12     & 41.82\mp0.13    \nl
log \fsixty/\fhundred    & $-$0.51\mp0.06   & $-$0.26\mp0.07   & $-$0.15\mp0.04  \nl
\Tdust\ (K)              & 29.9\mp1.4       & 36.4\mp2.3       & 40.2\mp1.4      \nl
\LIR/\LB                 & 0.61\mp0.20      & \lt0.59\mp0.15   & 6.90\mp0.90     \nl
\LHa/\LB                 & 0.0002\mp0.0001  & 0.0043\mp0.0016  & 0.0044\mp0.0012 \nl
\LIR/\MH2\ (\Lsun/\Msun) & \gt2.0\mp0.5     & \gt3.8\mp1.5     & 16.1\mp4.6      \nl
\LHa/\MH2\ (\Lsun/\Msun) & \gt0.001\mp0.001 & \gt0.021\mp0.006 & 0.012\mp0.005   \nl
\MHI/\LB\ (\Msun/\Lsun)  & 0.18\mp0.08      & 0.56\mp0.20      & 0.32\mp0.16     \nl
\MH2/\LB\ (\Msun/\Lsun)  & \lt0.43\mp0.10   & \lt0.20\mp0.06   & 0.67\mp0.20     \nl
\MH2/\MHI                & \lt1.70\mp0.91   & \lt0.29\mp0.04   & 3.44\mp0.80     \nl
\enddata
\tablenotetext{a}{Quantities involving IRAS measurements are global values for each interacting system.
All others use values for individual galaxies, when available.}
\end{deluxetable}

\begin{deluxetable}{lllll}
%\tablenum{7}
\tablecolumns{5}
\tablecaption{Interaction strength classifications.\label{tbl_class}}
\tablewidth{0pt}
\tablehead{\colhead{Class 1} & \colhead{Class 2} &\colhead{Class 3} &
\colhead{Class 4} & \colhead{Class 5} }
\startdata
UGC 5617/20 & Arp 248     & Arp 256    & NGC 4038/9    & NGC 1614 \nl
UGC 7776/7  & Arp 250     & UGC 813/6  & NGC 7592      & UGC 966  \nl
            & UGC 480     & UGC 993    & UGC 1720      & UGC 4509 \nl
            & UGC 2320    & UGC 2992   & UGC 3737      & UGC 8387 \nl
            & UGC 3031    & UGC 6224   & UGC 4757      &          \nl
            & UGC 8528/9  & UGC 7938/9 & UGC 4881      &          \nl
            & UGC 11673   & UGC 8135   & UGC 6471/2    &          \nl
            & UGC 12914/5 & UGC 8335   & UGC 10267     &          \nl
            &             & UGC 8641/5 & UGC 12699/700 &          \nl
            &             & UGC 10923  &               &          \nl
            &             & UGC 11175  &               &          \nl
            &             & UGC 11284  &               &          \nl
\enddata
\end{deluxetable}

%\begin{table}
%\dummytable\label{tbl_clsprop}
%\end{table}

\begin{deluxetable}{lccccc}
%\tablenum{8}
\tablecolumns{6}
\tablecaption{Mean properties of interaction class subsamples.\label{tbl_clsprop}}
\tablewidth{0pt}
\tablehead{
\colhead{Quantity\tablenotemark{a}} & \colhead{Class 1} & \colhead{Class 2} & \colhead{Class 3}
 &\colhead{Class 4} &\colhead{Class 5} }
\startdata
Number of systems        &  2              &  8              &  12             &  9              &   4             \nl
\LB\ (10$^9$ \Lsun)      & 33.3\mp21.6     & 56.0\mp21.1     & 51.6\mp14.0     & 38.5\mp9.7      & 22.9\mp1.5      \nl
\LIR\ (10$^9$ \Lsun)     & 19.1\mp22.0     & 35.5\mp9.6      & 101.8\mp24.3    & 113.9\mp39.1    & 161.9\mp44.5    \nl
\MHI\ (10$^9$ \Msun)     & 4.2\mp3.6       & 6.3\mp1.4       & 11.4\mp2.8      &  9.1\mp2.3      &  4.5\mp1.2      \nl
\MH2\ (10$^9$ \Msun)     & 7.9\mp9.5       & 17.8\mp6.6      & 17.6\mp3.7      & 20.1\mp7.5      & 14.0\mp3.0      \nl
log \LHa\ (ergs/s)       & 41.37\mp0.63    & 41.37\mp0.37    & 41.75\mp0.16    & 41.83\mp0.20    & 41.48\mp0.30    \nl
log \fsixty/\fhundred    & $-$0.44\mp0.16  & $-$0.46\mp0.05  & $-$0.25\mp0.03  & $-$0.23\mp0.04  & $-$0.17\mp0.06  \nl
\Tdust\ (K)              & 31.4\mp3.3      & 31.0\mp1.1      & 36.7\mp0.9      & 37.3\mp1.3      & 39.5\mp2.1      \nl
\LIR/\LB                 & 0.45\mp0.37     & 0.69\mp0.12     & 2.3\mp0.7       & 2.5\mp0.7       & 7.0\mp1.8       \nl
\LHa/\LB                 & 0.0016\mp0.0011 & 0.0010\mp0.0006 & 0.0040\mp0.0008 & 0.0042\mp0.0007 & 0.0032\mp0.0018 \nl
\LIR/\MH2\ (\Lsun/\Msun) & 2.7\mp0.4       & 3.4\mp0.9       & 9.3\mp3.5       & 8.6\mp1.9       & 12.5\mp3.9      \nl
\LHa/\MH2\ (\Lsun/\Msun) & 0.010\mp0.003   & 0.008\mp0.007   & 0.012\mp0.003   & 0.021\mp0.007   & 0.006\mp0.004   \nl
\MHI/\LB\ (\Msun/\Lsun)  & 0.12\mp0.04     & 0.18\mp0.04     & 0.28\mp0.10     & 0.35\mp0.09     & 0.20\mp0.06     \nl
\MH2/\LB\ (\Msun/\Lsun)  & 0.18\mp0.17     & 0.27\mp0.08     & 0.37\mp0.11     & 0.44\mp0.17     & 0.61\mp0.13     \nl
\MH2/\MHI                & 1.44\mp1.04     & 2.38\mp1.56     & 1.93\mp0.44     & 2.41\mp0.92     & 3.39\mp0.51     \nl
\enddata
\tablenotetext{a}{All quantities have been computed using global values for each interacting system.}
\end{deluxetable}

\clearpage

\begin{figure}
\begin{center}
\epsfig{file=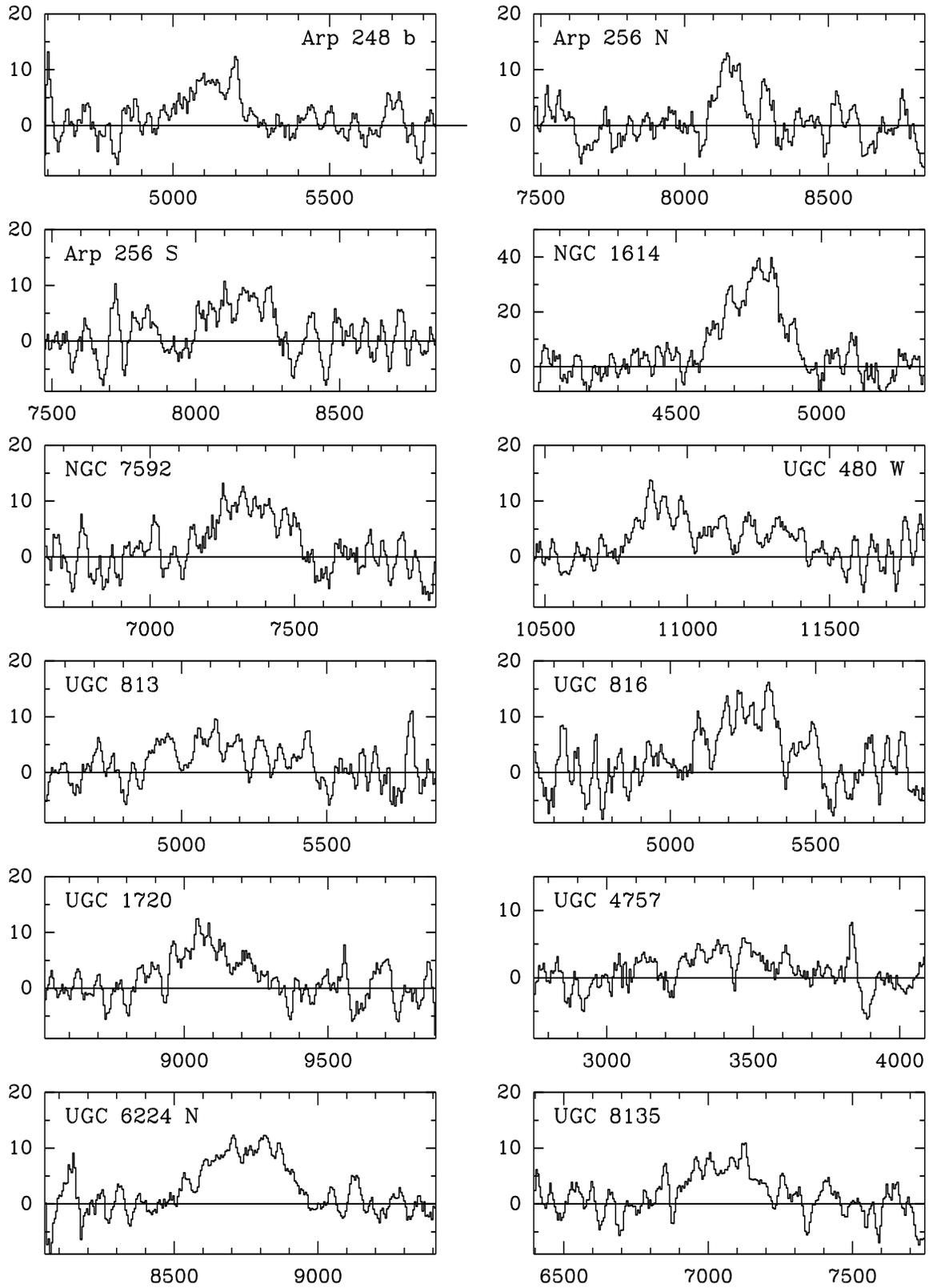,width=6.0in,angle=0}
\end{center}
\caption{CO spectra of the interacting galaxies.
The intensity scale is in units of $T_{R}^{*}$ (mK).
The $x$-axis is redshift expressed in units of \kms ($cz$).
\label{fig_cospectra}}
\end{figure}

\begin{figure}
\begin{center}
\epsfig{file=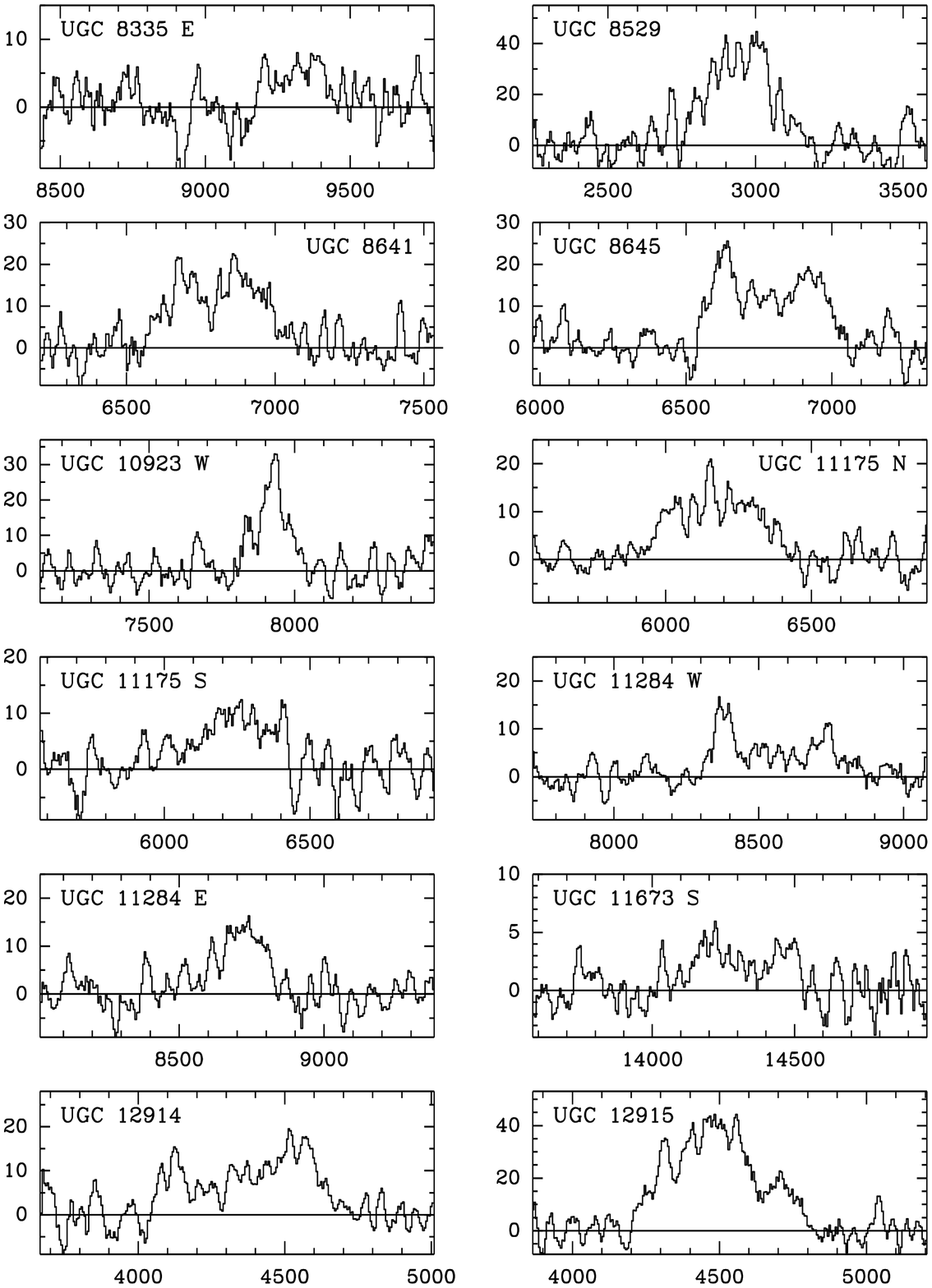,width=6.0in,angle=0}
\end{center}
\setcounter{figure}{0}
\caption{(continued)}
\end{figure}

\end{document}

%\ptlandscape

\setcounter{page}{14}

\begin{deluxetable}{lccccccccccc}
\tablenum{2}
\tablecolumns{12}
\tablecaption{Observed data for interacting systems. \label{tbl_obsdata}}
\tablewidth{596pt}
\tablehead{
\colhead{\ }   & \colhead{V$_0$}   & \colhead{\Bt}  & \colhead{\Ab}  &
\colhead{\SCO} & \colhead{\fsixty} & \colhead{\fhundred} & \colhead{log \FIR} &
\colhead{\SHI} & \colhead{\HI} & \colhead{log \FHa} & \colhead{\Ha} \\
\colhead{Object}&\colhead{(km/s)} & \colhead{(mag)} & \colhead{(mag)} &
\colhead{(Jy km/s)} & \colhead{(Jy)} & \colhead{(Jy)} & \colhead{(W/m$^2$)} &
\colhead{(Jy km/s)} & \colhead{Ref} & \colhead{(erg/cm$^2$/s)} & \colhead{Ref}\\
\colhead{(1)} & \colhead{(2)} & \colhead{(3)} & \colhead{(4)} & \colhead{(5)} &
\colhead{(6)} & \colhead{(7)} & \colhead{(8)} & \colhead{(9)} & \colhead{(10)} &
\colhead{(11)} & \colhead{(12)} }
\startdata
Arp 248a   & 5139    &14.50&0.04&\lt42\phd&  2.59&   5.41    &   $-12.82$ &\phn2.71 & 1 &$-$12.80&1b\nl
Arp 248b   & 4993    &14.10&0.04& \phn67  &      &           &            &\phn5.37 & 1 &$-$12.36&1b\nl
\tablevspace{0.1in}
Arp 250 W  &24181\phn&16.42&0.17&\lt32\phd&  0.21&   0.67    &   $-13.82$ &\nodata  &\nodata&\nodata&\nodata\nl
\tablevspace{0.1in}
Arp 256 N  & 8219    &14.81&0.11& \phn38  &  6.70&   11.1    &   $-12.45$ &\phn6.01 & 1 &$-$12.60&1a\nl
Arp 256 S  & 8215    &13.60&0.11& \phn63  &      &           &            &         &   &$-$12.38&1a\nl
\tablevspace{0.1in}
NGC 1614   & 4676    &13.63&0.22&  274    &  32.4&   39.0    &   $-11.81$ &\phn3.05 & 1 &$-$11.84&2\nl
\tablevspace{0.1in}
NGC 4038   & 1382    &10.91&0.05& 1150\phn&  43.5&   75.0    &   $-$11.63 &   37.42 & 2 &$-$10.79&2\nl
NGC 4039   & 1414    &11.08&0.05&  920    &      &           &            &         &   &      & \nl
\tablevspace{0.1in}
NGC 7592   & 7452    &14.22&0.10&  105    &  6.92&   12.2    &   $-$12.42 &\phn5.42 & 1 &$-$12.27&1a\nl
\tablevspace{0.1in}
UGC 480 W  &11228\phn&13.50&0.17&  131    &  1.40&\phn4.4    &   $-$12.99 &\lt1.93\phd& 1 &$-$13.08&1a\nl
UGC 480 E  &11302\phn&15.60&0.17&\lt32\phd&      &           &            &         &   &$-$15.00&1a\nl
\tablevspace{0.1in}
UGC 813    & 5316    &14.80&0.49& \phn68  &  2.76&   7.58    &   $-$12.73 &  10.50  & 1 &$-$12.66&1a\nl
UGC 816    & 5449    &14.20&0.49&  118    &      &           &            &         &   &$-$12.52&1a\nl
\tablevspace{0.1in}
UGC 966    & 2376    &12.24&0.05& 1260\phn&  31.1&   51.7    &   $-$11.78 &  27.53  & 2 &$-$12.40&2\nl
\tablevspace{0.1in}
UGC 993    & 2958    &15.60&0.14&\lt32\phd&  0.17&\lt0.31\phs&\lt$-$14.03\phs&\nodata&\nodata&$-$12.63&1b\nl
\tablevspace{0.1in}
UGC 1720   & 9120    &14.70&0.11& \phn88  &  5.26&   9.59    &   $-$12.53 &\phn2.96 & 1 &$-$12.38&1a\nl
\tablevspace{0.1in}
UGC 2320 SW&10432\phn&15.20&0.36&\lt42\phd&  0.15&\lt0.78\phs&\lt$-$13.83\phs&   \phn1.62&1&$-$14.45&1a\nl
\tablevspace{0.1in}
UGC 2992   & 4983    &15.40&0.48&\lt21\phd&  1.03&   1.33    &   $-$13.30 &\phn2.80 & 1 &$-$12.71&1b\nl
\tablevspace{0.1in}
UGC 3031   & 4632    &14.90&0.25&\lt39\phd&  0.35&   0.93    &   $-$13.64 &\phn2.42 & 1 &\nodata&\nodata\nl
\tablebreak
UGC 3737   & 4379    &15.30&0.32&\lt32\phd&  0.45&   1.18    &   $-$13.53 &\phn5.95 & 1 &$-$12.84&1b\nl
\tablevspace{0.1in}
UGC 4509   & 5248    &13.99&0.09&  170    &  24.0&   29.2    &   $-$11.94 &\phn2.11 & 1 &$-$12.77&3\nl
\tablevspace{0.1in}
UGC 4718 N & 3133    &13.25&0.08& \phn47  &  1.88&   3.03    &   $-$13.00 &  18.22  & 2 &$-$12.16&1b\nl
\tablevspace{0.1in}
UGC 4744 W & 2385    &14.04&0.10& \phn35  &  1.06&   3.52    &   $-$13.10 &  15.90  & 1 &\nodata&\nodata\nl
\tablevspace{0.1in}
UGC 4757   & 3342    &13.90&0.08&\lt46\phd&  1.45&   2.59    &   $-$13.10 &\phn7.58 & 1 &$-$12.74&1b\nl
\tablevspace{0.1in}
UGC 4881   &11959\phn&14.90&0.00&  200    &  6.18&   10.8    &   $-$12.47 &\phn3.26 & 1 &$-$13.17&4\nl
\tablevspace{0.1in}
UGC 5617   & 1278    &12.30&0.02&\lt90\phd&  7.84&   16.9    &   $-$12.33 &\phn6.79 & 3 &$-$13.52&1b\nl
UGC 5620   & 1028    &11.10&0.02&  450    &      &           &            &   23.14 & 2 &$-$11.74&1b\nl
\tablevspace{0.1in}
UGC 6224 N & 8710    &14.40&0.00&  125    &  4.06&   8.17    &   $-$12.63 &\phn1.65 & 1 &\nodata&\nodata\nl
UGC 6224 S & 8699    &14.70&0.00&\lt49\phd&      &           &            &         &   &\nodata&\nodata\nl
\tablevspace{0.1in}
UGC 6471   & 3237    &11.80&0.00&  610    &  114.&   129.    &   $-$11.27 &\phn5.90 & 1 &$-$11.09&2\nl
UGC 6472   & 3011    &11.98&0.00&  290    &      &           &            &         &   &      & \nl
\tablevspace{0.1in}
UGC 7776   & 2168    &11.68&0.00& 1050\phn&  16.7&   59.3    &   $-$11.89 &   10.98 & 2 &$-$11.54&2\nl
UGC 7777   & 2230    &12.06&0.00&  500    &      &           &            &   22.32 & 2 &$-$11.89&2\nl
\tablevspace{0.1in}
UGC 7938   & 6652    &14.70&0.01& \phn71  &  2.76&   5.41    &   $-$12.80 &\phn1.88 & 1 &$-$13.18&1b\nl
UGC 7939   & 6582    &14.40&0.01& \phn51  &      &           &            &\phn1.54 & 1 &$-$13.06&1b\nl
\tablevspace{0.1in}
UGC 8135   & 7074    &14.20&0.01& \phn74  &  6.17&   8.23    &   $-$12.52 &\phn0.70 & 3 &\nodata&\nodata\nl
\tablevspace{0.1in}
UGC 8335 W & 9233    &15.00&0.03&\lt32\phd&  11.4&   14.1    &   $-$12.26 &\phn1.20 & 3 &$-$13.14&1b\nl
UGC 8335 E & 9433    &15.00&0.03& \phn41  &      &           &            &         &   &$-$12.49&1b\nl
\tablevspace{0.1in}
UGC 8387   & 7000    &14.40&0.00&  220    &  15.2&   28.8    &   $-$12.07 &\phn2.95 & 4 &$-$12.55&5\nl
\tablevspace{0.1in}
UGC 8528   & 3039    &13.56&0.03&\lt95\phd&  7.02&   15.5    &   $-$12.37 &\phn4.31 & 5 &$-$13.60&1b\nl
UGC 8529   & 3067    &13.10&0.03&  350    &      &           &            &\phn3.88 & 5 &$-$13.31&1b\nl
\tablebreak
UGC 8641   & 6761    &12.87&0.01&  206    &  11.0&   21.5    &   $-$12.20 &\phn6.37 & 1 &$-$11.97&1a\nl
UGC 8645   & 6727    &12.93&0.01&  240    &      &           &            &\phn5.36 & 1 &$-$12.54&1a\nl
\tablevspace{0.1in}
UGC 10267  & 9032    &13.97&0.00&  200    &  6.27&   10.6    &   $-$12.47 &\phn5.29 & 1 &$-$12.06&1b\nl
\tablevspace{0.1in}
UGC 10923 W& 8098    &14.30&0.21&  113    &  4.65&   11.7    &   $-$12.53 &\phn4.33 & 1 &$-$12.32&1a\nl
UGC 10923 E& 7993    &16.31&0.21&\lt32\phd&      &           &            &         &   &$-$13.15&1a\nl
\tablevspace{0.1in}
UGC 11175 N& 6369    &14.00&0.22&  170    &  6.78&   13.8    &   $-$12.40 &\phn6.58 & 1 &$-$12.49&1a\nl
UGC 11175 S& 6436    &16.00&0.22&  103    &      &           &            &         &   &$-$13.52&1a\nl
\tablevspace{0.1in}
UGC 11284 W& 8757    &15.80&0.19&  116    &  9.20&   14.3    &   $-$12.32 &\phn3.56 & 1 &$-$12.72&1a\nl
UGC 11284 E& 8904    &15.70&0.19&  102    &      &           &            &\phn5.66 & 1 &$-$12.56&1a\nl
\tablevspace{0.1in}
UGC 11673 N&14428\phn&16.66&0.25&\lt32\phd&  0.80&   2.15    &   $-$13.27 &\nodata  &\nodata&\nodata&\nodata\nl
UGC 11673 S&14450\phn&15.68&0.25& \phn42  &      &           &            &\nodata  &\nodata&\nodata&\nodata\nl
\tablevspace{0.1in}
UGC 12699  & 2983    &13.00&0.16&  130    &  10.0&   12.7    &   $-$12.31 &   21.00 & 1 &$-$11.66&1b\nl
UGC 12700  & 2895    &14.68&0.16&\lt12\phd&      &           &            &         &   &$-$14.00&1b\nl
\tablevspace{0.1in}
UGC 12914  & 4555    &13.07&0.15&  213    &  5.77&   15.0    &   $-$12.42 &\phn7.00 & 1 &$-$12.53&1a\nl
UGC 12915  & 4653    &13.95&0.15&  502    &      &           &            &         &   &$-$12.73&1a\nl
\enddata
\tablecomments{
Col.~(1) Galaxy name.
Col.~(2) Galactocentric velocity, assuming
$V_0=V_{\sun}+9cos(l)cos(b)+232sin(l)cos(b)+7sin(b)$ (De Vaucouleurs
et~al.~1991: RC3).
Col.~(3) Total blue magnitude from the RC3 or UGC catalogs.
Col.~(4) Galactic extinction from Burstein \& Heiles 1982.
Col.~(5) CO flux from Tables~1 and 2.
Cols.~(6)--(7)  \IRAS\ flux densities at 60 and 100 \um. All
measurements are totals for each system and are listed only for the first
object in a pair.
Col.~(8) Far-infrared flux computed according to
$F_{\rm IR}=1.26(2.58\times 10^{-14}\fsixty + 1\times 10^{-14}\fhundred)$,
using the 60 and 100\um\ fluxes from columns 6 and 7.
Col.~(9) \HI\ flux. Measurements that are totals for a system are listed
only for the first object in a pair.
Col.~(10) Sources of the \HI\ data: (1) Bushouse 1987, (2) RC3, (3)
Huchtmeier \& Richter 1989, (4) Ritchter, Sackett, \& Sparke 1994, (5)
Haynes \& Giovanelli 1991.
Col.~(11) \Ha\ flux. Measurements that are totals for a system are listed
only for the first object in a pair.
Col.~(14) Sources of the \Ha\ data: (1a) Bushouse 1987---narrow-band
imaging, (1b) Bushouse 1987---22\arcsec\ aperture spectrophotometry,
(2) Young et~al.~1996---narrow-band imaging, (3) Armus, Heckman, \& Miley
1990---narrow-band imaging, (4) Veilleux et~al.~1995---long-slit
spectrophotometry, (5) Kennicutt~1998---narrow-band imaging.
}
\end{deluxetable}

\clearpage

\begin{deluxetable}{lcccccccccccc}
\tablenum{3}
\tablecolumns{13}
\tablecaption{Derived data for interacting systems. \label{tbl_derdata}}
\tablewidth{677pt}
\tablehead{
 & & & & & & & & & & \colhead{log} && \\
\colhead{\ }   & \colhead{$D$}  & \colhead{\LB}  & \colhead{\LIR}  &
\colhead{\MH2} & \colhead{\MHI} & \colhead{$\frac{\LIR}{\LB}$} &
\colhead{$\frac{\LIR}{\MH2}$} & \colhead{$\frac{\MH2}{\LB}$}  &
\colhead{$\frac{\MH2}{\MHI}$} & \colhead{\LHa} & \colhead{$\frac{\LHa}{\LB}$} &
\colhead{$\frac{\LHa}{\MH2}$} \\
\colhead{Object} & \colhead{(Mpc)} & \colhead{(10$^9$ \Lsun)} &
\colhead{(10$^9$ \Lsun)} & \colhead{(10$^9$ \Msun)} & \colhead{(10$^9$ \Msun)} &
 & \colhead{($\frac{\Lsun}{\Msun}$)} & \colhead{($\frac{\Msun}{\Lsun}$)} & &
\colhead{(ergs/s)} & & \colhead{($\frac{\Lsun}{\Msun}$)} \\
\colhead{(1)} & \colhead{(2)} & \colhead{(3)} & \colhead{(4)} & \colhead{(5)} &
\colhead{(6)} & \colhead{(7)} & \colhead{(8)} & \colhead{(9)} & \colhead{(10)} &
\colhead{(11)} & \colhead{(12)} & \colhead{(13)} }
\startdata
Arp 248a   &    68 & 11.7    &    21.6 &\lt2.1\phd&\phn2.9&  0.8&  6.4&\lt0.18\pl&\lt0.72\pl&40.95 & 0.0020 &\gt0.011\pl\nl
Arp 248b   &       & 16.9    &         &\phn3.4& \phn5.8 &      &     &   0.20& 0.58&41.39 & 0.0038 &   0.019 \nl
\tablevspace{0.1in}
Arp 250 W  &322\phn& 51.2    &    49.2 &\lt36.6\phs&\nodata&1.0 &\gt1.3\pl&\lt0.72\pl&\nodata&\nodata &\nodata &\nodata  \nl
\tablevspace{0.1in}
Arp 256 N  &110\phn& 24.6    &133.1\phn&\phn5.0&    17.0 &   1.3&10.0\phn&0.20& 0.78&41.58 & 0.0041 &   0.020 \nl
Arp 256 S  &       & 75.0    &         &\phn8.3&         &      &      &   0.11&    &41.80 & 0.0022 &   0.020 \nl
\tablevspace{0.1in}
NGC 1614   &    62 & 26.2    &188.2\phn&   11.7& \phn2.8 &   7.2&16.1\phn&   0.45& 4.19&41.88 & 0.0076 &   0.017 \nl
\tablevspace{0.1in}
NGC 4038   &    19 & 24.5    &    25.5 &\phn4.4& \phn3.1 &   0.6&   3.2&   0.18& 2.58&41.84 & 0.0074 &   0.041 \nl
NGC 4039   &       & 20.9    &         &\phn3.5&         &      &      &   0.17&     &      &        &         \nl
\tablevspace{0.1in}
NGC 7592   &    99 & 34.6    &117.3\phn&   11.4&    12.6 &   3.4&10.3\phn&   0.33& 0.90&41.83 & 0.0051 &   0.015 \nl
\tablevspace{0.1in}
UGC 480 W  &150\phn&163.4\phn&    72.2 &   32.5&\lt10.3\pl&  0.4&   2.2&   0.20&\gt3.16\pl&41.39 & 0.0004 &   0.002 \nl
UGC 480 E  &       & 23.6    &         &\lt7.9\phd&      &      &      &\lt0.34\pl&     &39.47 &\lt0.0001\pl&\gt0.001\pl\nl
\tablevspace{0.1in}
UGC 813    &    72 & 15.1    &    30.0 &\phn3.9&    12.8 &   0.7&   2.8&   0.26& 0.83&41.25 & 0.0030 &   0.012 \nl
UGC 816    &       & 26.3    &         &\phn6.7&         &      &      &   0.25&     &41.39 & 0.0024 &   0.010 \nl
\tablevspace{0.1in}
UGC 966    &    32 & 20.8    &    52.1 &   13.9& \phn6.5 &   2.5&   3.7&   0.67& 2.13&40.69 & 0.0006 &   0.001 \nl
\tablevspace{0.1in}
UGC 993    &    39 &\phn1.6&\lt0.5\phd&\lt0.6\phd&\nodata&\lt0.3\pl&$\sim$0.8\pl&\lt0.35\pl&\nodata&40.67 & 0.0078 &\gt0.023\pl\nl
\tablevspace{0.1in}
UGC 1720   &122\phn& 33.6    &136.4\phn&   14.3&    10.3 &   4.1&   9.5&   0.43& 1.39&41.89 & 0.0061 &   0.014 \nl
\tablevspace{0.1in}
UGC 2320 SW&139\phn& 34.9  &\lt8.9\phd&\lt8.9\phd&\phn7.4&\lt0.3\pl&$\sim$1.0\pl&\lt0.26\pl&\lt1.21\pl&40.00 & 0.0001 &\gt0.001\pl\nl
\tablevspace{0.1in}
UGC 2992   &    66 &\phn7.4  & \phn6.9 &\lt1.0\phd&\phn2.9&0.9  &\gt6.8\pl&\lt0.14\pl&\lt0.35\pl&41.13 & 0.0047 &\gt0.034\pl\nl
\tablevspace{0.1in}
UGC 3031   &    62 &\phn8.2  & \phn2.7 &\lt1.6\phd&\phn2.2&0.3  &\gt1.7\pl&\lt0.20\pl&\lt0.75\pl&\nodata &\nodata &\nodata \nl
\tablebreak
UGC 3737   &    58 &\phn5.4  & \phn3.1 &\lt1.2\phd&\phn4.8&0.6  &\gt2.6\pl&\lt0.22\pl&\lt0.25\pl&40.85 & 0.0034 &\gt0.015\pl\nl
\tablevspace{0.1in}
UGC 4509   &    70 & 21.0    &175.7\phn&\phn9.2& \phn2.4 &   8.4&  19.2\phn&   0.44& 3.75&41.02 & 0.0013 &   0.003 \nl
\tablevspace{0.1in}
UGC 4718 N &    42 & 14.7    & \phn5.5 &\phn0.9& \phn7.5 &   0.4&   6.1&   0.06& 0.12&41.18 & 0.0027 &   0.044 \nl
\tablevspace{0.1in}
UGC 4744 W &    32 &\phn4.2  & \phn2.5 &\phn0.4& \phn3.8 &   0.6&   6.5&   0.09& 0.10&\nodata &\nodata &\nodata \nl
\tablevspace{0.1in}
UGC 4757   &    45 &\phn9.2  & \phn4.9 &\lt1.0\phd&\phn3.6&0.5  &\gt4.9\pl&\lt0.11\pl&\lt0.28\pl&40.65 & 0.0013 &\gt0.012\pl\nl
\tablevspace{0.1in}
UGC 4881   &160\phn& 43.4    &269.3\phn&   55.9&    19.6 &   6.2&   4.8&   1.29& 2.86&41.31 & 0.0012 &   0.001 \nl
\tablevspace{0.1in}
UGC 5617   &    15 &\phn4.5  & \phn3.5 &\lt0.2\phd&\phn0.4&0.2  &   3.0&\lt0.05\pl&\lt0.61\pl&38.94 & 0.0001 &\gt0.001\pl\nl
UGC 5620   &       & 13.6    &         &\phn1.2& \phn1.3 &      &      &   0.09& 0.91&40.72 & 0.0010 &   0.012 \nl
\tablevspace{0.1in}
UGC 6224 N &116\phn& 36.4    &    98.7 &   18.5& \phn5.3 &   1.5&   5.3&   0.51& 3.53&\nodata &\nodata &\nodata \nl
UGC 6224 S &       & 27.6    &         &\lt7.3\phd&      &      &      &\lt0.26\pl&     &\nodata &\nodata &\nodata \nl
\tablevspace{0.1in}
UGC 6471   &    42 & 51.4    &291.2\phn&   11.6& \phn2.4 &   3.1&17.0\phn&   0.23& 7.10&42.23 & 0.0086 &   0.038 \nl
UGC 6472   &       & 43.6    &         &\phn5.5&         &      &      &   0.13&     &      &        &         \nl
\tablevspace{0.1in}
UGC 7776   &    29 & 28.5    &    34.6 &\phn9.9& \phn2.2 &   0.7&   2.4&   0.35& 4.45&41.47 & 0.0029 &   0.008 \nl
UGC 7777   &       & 20.1    &         &\phn4.7& \phn4.5 &      &      &   0.24& 1.04&41.12 & 0.0014 &   0.006 \nl
\tablevspace{0.1in}
UGC 7938   &    88 & 16.1    &    38.6 &\phn6.1& \phn3.5 &   1.0&   3.7&   0.38& 1.76&40.79 & 0.0010 &   0.003 \nl
UGC 7939   &       & 21.2    &         &\phn4.4& \phn2.8 &      &      &   0.21& 1.54&40.91 & 0.0010 &   0.005 \nl
\tablevspace{0.1in}
UGC 8135   &    94 & 29.2    &    84.0 &\phn7.2& \phn1.5 &   2.9&  11.6\phn&   0.25& 4.93&\nodata &\nodata &\nodata \nl
\tablevspace{0.1in}
UGC 8335 W &124\phn& 24.8    &266.0\phn&\lt5.5\phd&\phn4.4&  5.4&  38.1\phn&\lt0.22\pl& 1.59&41.13 & 0.0014 &\gt0.007\pl\nl
UGC 8335 E &       & 24.8    &         &\phn7.0&         &      &      &   0.28&     &41.78 & 0.0064 &   0.023 \nl
\tablevspace{0.1in}
UGC 8387   &    93 & 23.6    &231.7\phn&   21.1& \phn6.1 &   9.8&  11.0\phn&   0.90& 3.48&41.47 & 0.0033 &   0.004 \nl
\tablevspace{0.1in}
UGC 8528   &    41 & 10.0    &    22.1 &\lt1.7\phd&\phn1.7&  0.9&   3.5&\lt0.17\pl&\lt1.02\pl&39.70 & 0.0001 &\gt0.001\pl\nl
UGC 8529   &       & 15.3    &         &\phn6.4& \phn1.5 &      &      &   0.42& 4.20&39.99 & 0.0002 &\lt0.001\pl\nl
\tablebreak
UGC 8641   &    90 & 90.3\phn&159.4\phn&   18.3&    12.2 &   0.9&   4.0&   0.20& 1.51&42.02 & 0.0030 &   0.015 \nl
UGC 8645   &       & 85.5\phn&         &   21.4&    10.2 &      &      &   0.25& 2.09&41.45 & 0.0009 &   0.003 \nl
\tablevspace{0.1in}
UGC 10267  &120\phn& 58.3    &153.6\phn&   31.9&    18.1 &   2.6&   4.8&   0.55& 1.76&42.18 & 0.0068 &   0.012 \nl
\tablevspace{0.1in}
UGC 10923 W&107\phn& 41.4    &106.2\phn&   14.3&    11.8 &   2.2&   7.4&   0.35& 1.22&41.87 & 0.0047 &   0.014 \nl
UGC 10923 E&       &\phn6.5  &         &\lt4.1\phd&      &      &      &\lt0.62\pl&     &41.04 & 0.0044 &\gt0.007\pl\nl
\tablevspace{0.1in}
UGC 11175 N&    85 & 34.9    &    90.7 &   13.6&    11.3 &   2.2&   4.1&   0.39& 1.93&41.50 & 0.0024 &   0.006 \nl
UGC 11175 S&       &\phn5.5  &         &\phn8.3&         &      &      &   1.49&     &40.47 & 0.0014 &   0.001 \nl
\tablevspace{0.1in}
UGC 11284 W&118\phn& 12.3    &207.4\phn&   17.7&    11.7 &   8.1&   6.2&   1.44& 1.52&41.54 & 0.0074 &   0.005 \nl
UGC 11284 E&       & 13.5    &         &   15.6&    18.5 &      &      &   1.15& 0.84&41.70 & 0.0098 &   0.009 \nl
\tablevspace{0.1in}
UGC 11673 N&193\phn& 15.7    &    62.2 &\lt13.1\phs&\nodata& 1.1&   3.6&\lt0.83\pl&\nodata&\nodata &\nodata &\nodata \nl
UGC 11673 S&       & 38.8    &         &   17.1& \nodata &      &      &   0.44&     &\nodata &\nodata &\nodata \nl
\tablevspace{0.1in}
UGC 12699  &    39 & 17.5    &    23.5 &\phn2.2& \phn7.6 &   1.1&  10.7\phn&   0.13& 0.29&41.64 & 0.0066 &   0.052 \nl
UGC 12700  &       &\phn3.7  &         &\lt0.2\phd&      &      &      &\lt0.06\pl&     &39.30 & 0.0001 &\gt0.003\pl\nl
\tablevspace{0.1in}
UGC 12914  &    61 & 39.8    &    44.8 &\phn8.8& \phn6.2 &   0.8&   1.5&   0.22& 4.76&41.16 & 0.0009 &   0.004 \nl
UGC 12915  &       & 17.7    &         &   20.8&         &      &      &   1.18&     &40.96 & 0.0013 &   0.001 \nl
\enddata
\tablecomments{
Col.~(1) Galaxy name.
Col.~(2) Distance in Mpc, computed using the mean V$_0$ of each system
from Table~2, and assuming $H_{\rm 0}=75$ km s$^{-1}$ Mpc$^{-1}$.
Col.~(3) Blue luminosity, computed from values of \Bt\ in Table~2, corrected
for Galactic extinction, and assuming $M_{B_{\sun}}=+5.48$.
Col.~(4) Far-infrared luminosity, computed from values of $F_{\rm IR}$ in
Table~2. Measurements are totals for each system and are listed only
for the first object in a pair. This applies to all quantities involving
\IRAS\ far-infrared measurements.
Col.~(5) \H2\ mass, computed using $\MH2= 1.1\times 10^4D^2\SCO$.
Col.~(6) \HI\ mass, computed using $\MHI=2.36\times 10^5D^2\SHI$.
Measurements that are totals for a system are listed only for the first object
in a pair.
Col.~(7) Far-infrared to blue luminosity ratio.
Col.~(8) Far-infrared luminosity to \H2\ mass ratio.
Col.~(9) \H2\ mass to blue luminosity ratio.
Col.~(10) \H2\ to \HI mass ratio. Measurements that are averages for a system
are listed only for the first object in a pair. 
Col.~(11) Logarithm of the \Ha\ luminosity. Measurements that are totals for
a system are listed only for the first object in a pair.
Col.~(12) \Ha\ to blue luminosity ratio.
Col.~(13) \Ha\ luminosity to \H2\ mass ratio.
}
\end{deluxetable}